\documentclass{JHEP3} 
\usepackage{epsfig}
\epsfclipon

\newcommand{\nco}{\newcommand}
\nco{\beq}{\begin{equation}}
\nco{\eeq}{\end{equation}}
\nco{\beqa}{\begin{eqnarray}}
\nco{\eeqa}{\end{eqnarray}}
\nco{\lra}{\leftrightarrow}

\nco{\sss}{\scriptscriptstyle}
\nco{\vpp}{\vec p^{\,2}} 
\nco{\lsim}{\mbox{\raisebox{-.6ex}{~$\stackrel{<}{\sim}$~}}}
\nco{\gsim}{\mbox{\raisebox{-.6ex}{~$\stackrel{>}{\sim}$~}}}

\def\negsp{\!\!\!\!\!\!\!\!\!\!}
\def\m{\mu}
\def\l{\ell}
\def\diag{{\rm diag}}

\title{\Large Cosmology of codimension-two braneworlds}

\author{James M.~Cline$^{1,2}$, Julie Descheneau,$^2$, Massimo Giovannini$^1$, J\'er\'emie Vinet$^2$\\
$^1$ Theory Division, CERN CH-1211, Geneva 23, Switzerland\\
$^2$ Physics Department, McGill University,
3600 University Street, Montr\'eal, Qu\'ebec, Canada H3A 2T8\\
E-mail: \email{James.Cline@cern.ch}, \email{jdesch@hep.physics.mcgill.ca}, \email{Massimo.Giovannini@cern.ch},
\email{vinetj@physics.mcgill.ca}}
\preprint{CERN-TH/2003-089, McGill 03-07}

\abstract{We present a comprehensive study of the cosmological solutions of 6D braneworld
models with azimuthal symmetry in the extra dimensions, moduli stabilization by flux or a bulk
scalar field, and which contain at least one 3-brane that could be identified with our world. 
We emphasize an unusual property of these models: their expansion rate depends on the 3-brane
tension either not at all, or in a nonstandard way, at odds with the naive expected
dimensional reduction of these systems to 4D general relativity at low energies.  Unlike other
braneworld attempts to find a self-tuning solution to the cosmological constant problem, the
apparent failure of decoupling in these models is not associated with the presence of
unstabilized moduli; rather it is due to automatic cancellation of the brane tension by the
curvature induced by the brane. This provides some corroboration for the hope that these
models  provide a distinctive step toward understanding the smallness of the observed
cosmological constant.  However, we point out some challenges for obtaining realistic
cosmology within this framework. }

\begin{document}
\section{Introduction}\normalsize

The cosmology of codimension one branes, notably 3-branes in a 5D universe, has been
intensively studied and is now rather well understood.  One of the striking predictions
of this picture is that the Friedmann equation for the Hubble expansion rate should have
the form $H^2 = {8\pi\over 3} G\rho(1+{\rho\over T})$ in the simplest situation with a
single brane of tension $T$ and excess energy density $\rho$, in a warped space where
the bulk is 5D anti-deSitter space \cite{BDL}-\cite{CGS}.  One naturally wonders whether this kind of 
behavior is particular to branes of codimension one ({\it i.e.,} having only one
transverse dimension), or if higher codimension branes can act similarly.   There have
been numerous attempts over the last twenty years to harness some of the unusual
features of six dimensional models in order to get some insight into the cosmological
constant problem \cite{RuSh}-\cite{nav}, the weak scale hierarchy problem
\cite{Wett,wilt},\cite{CK}-\cite{BCCF} or other effects of warped compactification
\cite{Gio}-\cite{Multamaki}, and moduli stabilization
\cite{BCCF},\cite{Ponton}-\cite{GMZ}.  However we are not aware of any attempts to
systematically explore modifications to the Friedmann equation in codimension two
braneworld models. In fact, most of the work referred to above has been confined to the
study of static solutions.

In this paper we consider the most symmetric case, where the extra dimensions
have azimuthal symmetry and the metric functions depend only upon the radial coordinate
of the bulk, with periodic angular coordinate $\theta\in[0,2\pi]$:
\beq
\label{ansatz}
	ds^2 = M^2(\rho)g_{\mu\nu} dx^\mu dx^\nu + d\rho^2 + L^2(\rho) d\theta^2;
\eeq
here $g_{\mu\nu}$ is a maximally symmetric 4D metric which can be deSitter,
anti-deSitter, or Minkowski space, with cosmological constant $\Lambda_4$, normalized
such that the scale factor is $a(t) = e^{Ht}$  with $H=\kappa_6\sqrt{\Lambda_4/3}$ in
terms of the 6D gravitational constant $\kappa_6$.  General FRW solutions in the 4D
spacetime would require having distinct warp factors for $dt^2$ and $(dx^i)^2$, a
complication which we choose to avoid in the present study.  We will assume that the
only bulk contributions to the stress energy are a 6D cosmological constant $\Lambda_6$,
a magnetic flux $F_{\rho\theta}$, or a bulk scalar field $\phi$.

In section 2 we will classify all the possible bulk solutions which are consistent with
(\ref{ansatz}) and point out the insensitivity of $\Lambda_4$ to 3-brane tensions in
some of these solutions, including a particularly novel one in section 2.1. In section 3 we discuss
the boundary conditions associated with the tensions of branes which bound these
solutions, which are the source of any dependence of the Hubble rate on the brane
tensions which does exist. Section 4 sets the stage for investigating the Friedmann
equation in a  class of warped solutions by reviewing how these solutions can be 
stabilized by a bulk scalar field. This solution is known as the AdS soliton, since  it
looks like AdS$_6$ at large radii. In section 5 we derive a surprising result for the
AdS soliton, namely that the Friedmann equation relating the expansion rate to the brane
tensions has an unusual form which does not agree with the expectation that the system
should reduce to 4D general relativity at distances much larger than the
compactification scale.  In section 6 we turn our attention to an unwarped solution,
studied recently in \cite{CG,nav}, where the bulk is a two-sphere stabilized by magnetic
flux and deformed by deficit angles due to 3-branes at antipodal points. We look for
unstable modes in the spectrum of fluctuations around this solution which might explain
why the expansion rate of these models is insensitive to the 3-brane tensions, and show
that no such modes exist.  We conclude with a discussion of how 2D compactification
manifolds manage to tune away the effects of 3-brane tensions in certain cases, and the
difficulties that will have to be overcome if one wants to study realistic cosmology
by putting matter with a general equation of state on the branes.

\section{Bulk Solutions}
We start by writing the Einstein and scalar field equations for the metric
(\ref{ansatz}).  Defining $\m=M'/M$, $\l = L'/L$, and working in units where
the 6D gravitational constant $\kappa_6^2 = 1$, these  equations are
\beqa
\label{mumu}
\negsp{\mu\mu:}\quad \l'+3\m' + \l^2+6\m^2+3\l\m &=& -\Lambda_6 + {\Lambda_4\over M^2}
	-\frac12\left(\phi'^2 + m^2\phi^2 + {n^2\phi^2\over L^2} \right) - {\beta^2\over
2 M^8}\\
\label{thth}
\negsp{\theta\theta:}\quad 4\m' + 10\m^2 \phantom{AAAAAAAA;}
&=& -\Lambda_6 + {2\Lambda_4\over M^2}
	-\frac12\left(\phi'^2 + m^2\phi^2 - {n^2\phi^2\over L^2} \right) + {\beta^2\over
2 M^8}\\
\label{rr}
\negsp{\rho\rho:}\quad 4\m\l + 6\m^2 \phantom{AAAAAAAAA}
&=& -\Lambda_6 + {2\Lambda_4\over M^2}
	+\frac12\left(\phi'^2 - m^2\phi^2 - {n^2\phi^2\over L^2} \right) + {\beta^2\over
2 M^8}\\
\label{phi}
\negsp\phi:\quad \phi'' + (4\m+\l)\phi' \phantom{AAAAAA} &=& \ m^2\phi + {n^2\phi^2\over L^2}
\eeqa
The 4D cosmological constant is defined by $H^2 = \Lambda_4/3$ in the de Sitter brane 
case, where the 4D line element has the FRW form $ds^2 = -dt^2 + e^{2Ht}d\vec x^{\,2}$
We have allowed for the scalar field to be either real or complex with a winding
number $n$.  The constant $\beta$ is related to the magnetic field strength by
$F_{\rho\theta} = \beta L/M^4$, which can be seen to satisfy the Maxwell equation
$\partial_A(\sqrt{|G|} F^{AB}) = 0$.  For simplicity we have assumed that the bulk
scalar has no self interactions, so $V(\phi) = \frac12 m^2\phi^2$.  
Only three of these equations are independent; for example the $(\mu\mu)$ equation
can be derived from differentiating $(\rho\rho)$ and combining with the other equations.

Rubakov and Shaposhnikov \cite{RuSh} found an elegant way to solve the 6D Einstein
equations resulting from the metric (\ref{ansatz}), which was generalized to include
bulk magnetic flux in \cite{Wett} (see also \cite{wilt}).   
We extend the method
to take into account the possible presence of bulk scalar fields.
The Einstein equation (\ref{thth})
 for this system is equivalent to the equation governing
the classical motion of a particle of unit mass, whose action is
\beq
	S = \int d\rho \left(\frac12 z'^2 - U(z) -U_\phi(z)\right).
\eeq
Here $z' = {dz\over d\rho}$ and the potential depends on the 
4D and 6D cosmological
constants and the magnetic field via\footnote{we correct a 
factor of 2 error in $b$ relative to \cite{RuSh}}
\beqa
	U(z) &=& a z^2 - b z^{6/5} + c z^{-6/5};\quad
	a = \frac{5}{16}\Lambda_6;\quad b=\frac{25}{24}\Lambda_4;\quad c =
	\frac{25}{96}\beta^2 \\
	U_\phi(z) &=& \frac{5}{32}z^2\left(\phi'^2+m^2\phi^2-{n^2\phi^2\over L^2}\right)
\eeqa
The radial coordinate $\rho$ plays the role of time in this analogy.  If we ignore
the bulk scalar $\phi$, the resulting equation of motion $z'' + U'(z)=0$ allows
one to solve for the warp factor $M(\rho)$ independently of any other fields in the
problem, using the relation
\beq
\label{Mzrel}
	M(\rho) = z^{2/5}(\rho).
\eeq
If $\phi$ is nonzero, this is no longer possible; however one can find an approximate
solution iteratively by first finding $M(z)$ when $\phi=0$, next solving for $\phi$
in the background geometry, and then finding the approximate form of the back reaction 
of $\phi$ on the geometry by treating $U_\phi$ as a perturbation.

The above procedure has so far yielded only the metric component $M$.  To obtain
$L$, we can in general solve the first order $(\rho\rho)$ equation; but in most
cases there is a much simpler relation which gives either an exact expression or
a good approximation to $L$ in terms of $M$:
\beq
\label{Lrel}
	L(\rho) = R {dM\over d\rho}.
\eeq
Here $R$ is a constant of integration, which as we will show later
is determined by the tension of a 3-brane which may be located at $\rho=0$.
This result follows from the difference of the $(\theta\theta)$ and $(\rho\rho)$ 
equations, which can be written as
\beq
	{L'\over L} = {M''\over M'} + {\phi'^2\over 4\m}
\eeq
Thus the relation (\ref{Lrel}) is exact whenever $\phi'=0$, with the exception of
unwarped solutions where $M'=0$.  For these the $(\rho\rho)$ Einstein equation provides
no information about $L$, and we must turn to the ($\mu\mu$) equation, as will be 
discussed in section \ref{sect2.3}.   

Let us now focus on situations in which $\phi'$ is negligible and the scalar field has no
winding; in this case $U_\phi$ can be absorbed into the $a$ term of $U$. The potential $U(z)$
has eight possible distinctive shapes depending on the signs of $\mu$ and $\Lambda$, and
whether or not there is a magnetic flux. These are summarized in figure 1.  The solutions can
be visualized by starting the particle at rest ($z' = 0$) at position  $\rho=0$ with some
initial value $z_0$,  and letting it roll in the potential.  The position $\rho$ should be
thought of as time in the mechanics analogy. The initial condition that $z'=0$ implies that
$M'=0$ at $\rho=0$, which must be the case; there is no singular source term for $M''$ in the
Einstein equations, and therefore $M'$ must be continuous at the origin.  Considering $M'$
evaluated along a straight line passing through the origin, one sees that continuity of the
slope requires that $M'$ in fact vanishes. Using the mechanics analogy, a closed form solution
for $M(\rho)$ is implicitly  provided by integrating the equation of energy conservation, $E =
\frac12 z'^2+U(z)$: 
\beq
\label{rhoint}
	\rho = \left| \int_{z_0}^z {dz\over \sqrt{2(E - U(z))}} \right|
\eeq

\FIGURE{
\centerline{\epsfxsize=6.in\epsfbox{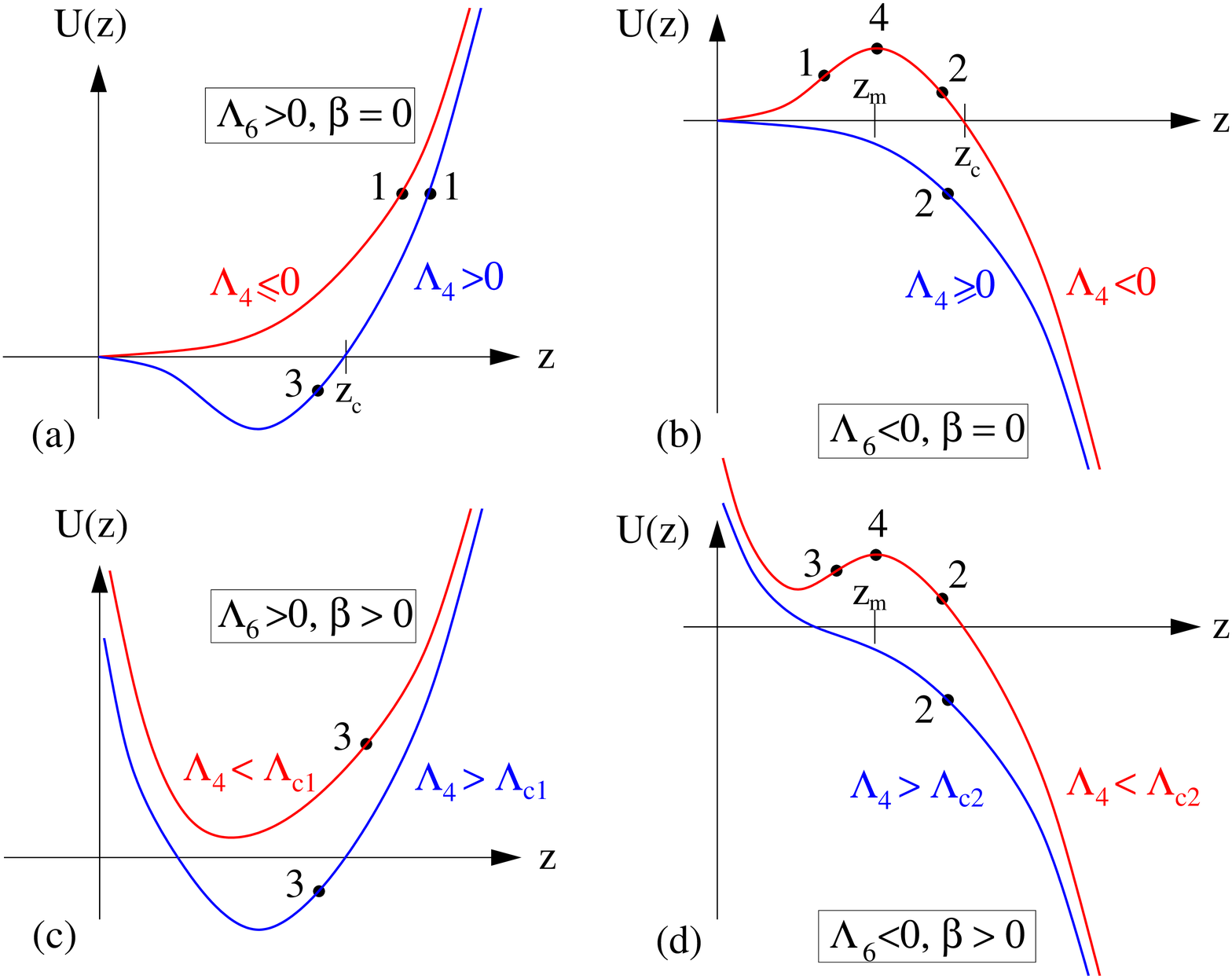}}
\caption{\small
Shapes of the potential $U(z)$ depending on the sign of $\Lambda_4$ for (a)
$\Lambda_6>0$ and (b) $\Lambda_6<0$
in the case of vanishing magnetic field $\beta$, and similarly in (c) and (d) for
nonzero $\beta$.
Numbered dots show initial conditions leading to the four kinds of solutions discussed below.
The first critical value is given by $\Lambda_{c1}^4 = (2\Lambda_6/5)^3\beta^2$. }
}

Once the particle starts rolling, there are four possible outcomes, which depend on the shape of the
potential and the initial value $z_0$. 
\begin{enumerate}

\item In one kind of solution, the particle can reach $z=0$ at some finite value of $\rho$. 
This happens when $\beta=0$, $\Lambda_6>0$ and $E>0$  (fig.\ 1a), or when $\beta=0$, 
$\Lambda_6<0,\
\Lambda_4<0$ and $z<z_m$ (fig.\ 1b).   Some curvature invariants diverge, so the extra
dimensions end at a singularity with the topology of $S_1$ in this case \cite{Gio}.  

\item Another kind is where $z\to\infty$ as $\rho\to\infty$, as occurs if $\Lambda_6<0$ and
$\Lambda_4\ge 0$, or if $\Lambda_6<0$, $\Lambda_4 < 0$ and $z>z_m$. Then gravity will not be localized,
and the world does not look 4-dimensional to an observer on the brane.  A special case of this
kind is the AdS soliton solution \cite{HM}, where $\beta=\Lambda_4=0$.

\item A third possibility is that the particle starts to the left of a stable
equilibrium position, in which case it comes to rest again at the other side of the
minimum at some maximum value of $\rho=\rho_m$.  The vanishing of $z'$ means that the
space closes off at $\rho_m$, and we can consistently insert a second 3-brane at this
position. The extra dimensions look like $S_2$ deformed by deficit angles due to the two
3-branes at antipodal points.  A special case \cite{CG,nav} is when the particle stays
at rest at the equilibrium position.  This gives an unwarped solution since $M(\rho)$
remains constant.  Nonvanishing magnetic flux $\beta$ is required to obtain static
solutions of this kind.

\item One can also find special unwarped solutions when the particle sits at an unstable
equilibrium position, denoted in fig.\ 1 by $z_m$, in the case of an AdS 3-brane.  Since
we are interested in cosmological solutions, we do not pursue these here.

\end{enumerate}

To cure the problems of cases (1) and (2) above, one should cut out the part of the
space with $z=0$ or $z=\infty$ at some value $z=z_4$, and insert a 4-brane at the
corresponding  radius  $\rho=\rho_m$ \cite{BCCF}. In fact it is not generally possible
to satisfy the jump conditions  (to be described in section 3) at  $\rho_m$ using an
ordinary 4-brane, since the bulk solution requires that its stress-energy components 
$T_{\theta\theta}$ differ from the other spatial components $T_{ii}$. One way of
accomplishing this is to imagine that a pure tension 4-brane is accompanied by a 3-brane
which is smeared around the compact extra dimension \cite{LMW}, or by Casimir energy of
a massless field confined to the 4-brane \cite{CN}.  Alternatively, it is possible to
find a special value of  $\rho_m$ where the jump conditions on $M$ and $L$ can both be
satisfied for a pure tension 4-brane, as long as $\Lambda_4$ is nonzero.  By integrating
the exact result (\ref{mleq}) which we will derive in section 4.3, this value is
implicitly given by
\beq
	\int_0^{\rho_m} M^2 L\, d\rho = {L'(0)\over \Lambda_4}
\eeq
showing that $\rho_m\to\infty$ as $\Lambda_4\to 0$ for the warped solutions where
$M$ and $L$ grow exponentially.  If one wants a solution which
remains compact in the static limit, this is not acceptable.

For completeness we mention one other kind of solution: we can start with $z'\neq 0$ at
$\rho=0$, but in this case the position $\rho=0$ should not be regarded as a single point, but
rather the location of another 4-brane.  A special case is when $\Lambda_6<0$, $\Lambda=E=0$;
then the warp factor takes the simple exponential form of the RS model.  In fact this is just
the 5D RS solution augmented by one extra compact dimension, whose warp factor is identical to
that of the large dimensions.  This model does not involve 3-branes, and is mathematically
quite similar to the 5D case which has been so well studied already.  A more thorough
examination of the cosmology of these solutions in presented in appendix B, and we
comment upon its qualitative differences relative to the codimension two case in
the conclusions.

In the following subsections we present some specific analytic solutions illustrating
the cases mentioned above.

\subsection{Type 1 solutions}
A solution which is borderline between type 1 and type 3 can be found by 
choosing $\Lambda_4,\Lambda_6 > 0$, $E=0$. Starting from $z=z_c=(b/a)^{5/4}$, the
critical value shown in fig.\ 1(a) defined by $U(z_c)=0$, the particle subsequently
rolls toward $z=0$.  The solution is
\beq
\label{soln1}
	M(\rho) = \cos\left(k\rho
	\right);\quad L(\rho) = Rk\sin\left(k\rho
	\right);\quad k = \sqrt{\Lambda_6\over 10} = H
\eeq
where we have used the freedom to rescale $x^\mu$ to make $M(0)=1$. Recall that
$H=\sqrt{\Lambda_4/3}$ is the Hubble constant, so that $\Lambda_4$ is determined by
\beq
\label{friedeq0}
	\framebox{$\displaystyle \Lambda_4 = \frac3{10}\Lambda_6$}
\eeq
Interestingly, no curvature invariants diverge at the point $\rho_m = \pi/2k$ where $M=0$ and
the space should be terminated.   In fact the Ricci scalar is constant, $R = 3\Lambda_6$, and
so are the Ricci tensor squared $R_{\alpha\beta}R^{\alpha\beta} = \frac32\Lambda_6^2$, and
Riemann tensor squared $R_{\alpha\beta\gamma\delta}R^{\alpha\beta \gamma\delta} =
\frac35\Lambda_6^2$. The Weyl tensor vanishes. This is analogous to the behavior as $y\to\infty$
in the 5D Randall-Sundrum solution \cite{RS} with the line element $ds^2 = e^{-2k|y|}(-dt^2 + d\vec
x^{\,2}) + dy^2$; even though the warp factor vanishes as $y\to\infty$ in this model, there is
no singularity.  Although  $M'(\rho_m)\neq 0$, yet $z' = 5/2 M^{3/2} M'$ vanishes at $\rho_m$,
so in the mechanics analogy, the ball rolls to the top of the hill and comes to rest again.  
Apparently the 4D part of the spacetime disappears at $\rho_m$, leaving one with just 2D
Euclidean space in the $\rho$-$\theta$ plane.

In the case of vanishing $\Lambda_4$ and positive $\Lambda_6$, the following solution can be 
found \cite{Luty}
\beq
\label{soln1b}
	M(\rho) = \cos\left(k\rho
	\right)^{2/5};\quad L(\rho) = Rk\sin\left(k\rho
	\right)^{2/5};\quad k = \sqrt{\frac58 \Lambda_6}.
\eeq
In this case however, in the mechanics analogy the particle does not come to rest at $z=0$, and 
there will be a curvature singularity at that point.  It is therefore necessary to insert a 
4-brane at $\rho < \rho_s$ to cut off the space before reaching the point where it becomes 
singular.

\subsection{Type 2 solutions}
It is interesting to consider solutions with $\Lambda_6<0$ since this gives
a warped geometry, where the hierarchy problem can be solved on the
3-brane {\it \`a la} Randall-Sundrum. 
We find a solution of type 2 when $\beta=E=0$ and $\Lambda_4, \Lambda_6 < 0$  by analytically
continuing (\ref{soln1}) in $\Lambda_4$ and $\Lambda_6$:
\beq
\label{soln2}
	M(\rho) = \cosh\left(k\rho
	\right);\quad L(\rho) = R k\sinh\left(k\rho
	\right);\quad k = \sqrt{|\Lambda_6|\over 10} = iH
\eeq
This follows the upper curve of fig.\ 1(b), starting at $z_c$. The
Hubble constant is imaginary because the 4D metric is AdS space.
{In addition the $E=0$ type 2 solution with $\Lambda_4>0$ and $\Lambda_6<0$ is
\[
	M(\rho) = \sinh\left(k\rho
	\right);\quad L(\rho) = Rk\cosh\left(k\rho
	\right)
\]
which follows the lower curve of fig.\ 1(b), starting at $z=0$. However, this solution does not admit a
3-brane at $\rho=0$ since $L$ is nonvanishing at this point.}

The only way to admit a 3-brane at $\rho=0$ if $\Lambda_4\ge 0$ is to take $E<0$, so
that the initial condition $z'=0$ (equivalent to the boundary condition $M'=0$) can be
satisfied. 
The static solution with $\Lambda_4=0$ (and $\beta=0$) has been
investigated in \cite{LMW, BCCF}; this is the AdS soliton truncated at large $\rho$ by a
4-brane:
\beq
\label{adssol}
	M(\rho) =  \cosh^{\frac25}(k\rho);\quad L(\rho) = \frac25 R k
 {\sinh(k\rho)\over\cosh^{\frac35}(k\rho)}\quad k = \sqrt{-\frac58 \Lambda_6}
\eeq

Analytic expressions for $M(\rho)$ in the AdS soliton type of solution no longer exist when
$\Lambda_4 > 0$.  However, we can obtain approximate analytic solutions by considering
$\Lambda_4$ as a perturbation.  The details are given in appendix A.  The result is that
the unperturbed solution gets shifted according to 
\beqa
\label{adssol2}
	M(\rho) &=& \left[ z_0 \cosh(k(\rho + \delta\rho))\right]^{2/5};
	\quad L(\rho) = R M'(\rho)
\eeqa
where $\delta\rho(z)$ has the shape shown in fig.\ 2, and $z = z_0\cosh(k\rho)$.
At sufficiently large values of $\rho$, the shift $\delta\rho$ approaches a constant,
so at leading order the only difference between the perturbed and unperturbed solutions at large
$\rho$ is a shift in the radial size of the extra dimension.

\FIGURE{
\centerline{\epsfxsize=5.in\epsfbox{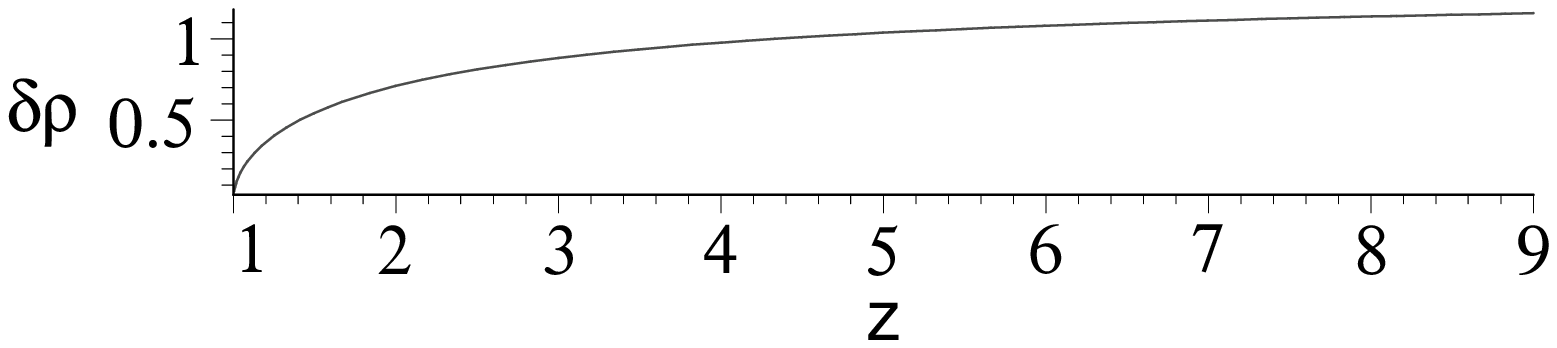}}
\caption{\small
$\delta\rho$ versus $z$, due to nonzero $\Lambda_4$ in the AdS soliton solution.}
}

\subsection{Type 3 solutions}
\label{sect2.3}
The simplest example of a solution of type 3 is that in which the particle sits at a minimum
of $U$.  In this case $M$ is constant, so the solution is unwarped.  If there is no magnetic
field, this occurs only in fig.\ 1(a), when $\Lambda_4>0$.  To specify the physical value
of $\Lambda_4$, we should rescale $M\to 1$ so that time and energy are normalized in the
usual way ($ds^2 =-dt^2 + e^{2Ht}d\vec x^{\,2}$).  Since $\Lambda_4 = 3H^2$, which has
dimensions of (mass)$^2$, the relevant quantity is $\Lambda_4/M^2 = \Lambda_6/2$; in other
words the expansion rate is governed completely by $\Lambda_6$, which is to be expected since
it is the only source of stress energy.

Another example of unwarped solutions is the static case $\Lambda_4=0$ with nonvanishing
magnetic field.  Just like above we noted that the physically relevant combination for the
expansion rate is  $\Lambda_4/M^2$, in this case $\beta^2/M^8$ is the meaningful combination
for the square of the magnetic field strength.  Minimizing $U(z)$ we find that $\beta^2/M^8 = 
2\Lambda_6$, in agreement with references \cite{CG,nav}.  Notice that $\Lambda_6$ must be
positive to achieve this fine tuning with the magnetic flux. 

In fact, the general relation between $\Lambda_4$, $\Lambda_6$ and $\beta$ can be easily
found in the same way, demanding $U'(z_0)=0$, when we realize that there is the freedom
to set $z_0\to 1$ by rescaling $x_\mu\to x_\mu/z_0$.  Thus 
\beq
\label{friedeq1}
	\framebox{$\displaystyle\Lambda_4 = \frac12\Lambda_6 - \frac14 \beta^2$}
\eeq
characterizes the physical expansion rate for an observer on the 3-brane when the bulk parameters 
are not tuned to give a static solution.  This is a significant result because it is completely
independent of the tension of such a 3-brane \cite{CG,nav}.

Although $M$ is trivial for the unwarped solutions, $L$ is nontrivial:
\beq
	L(\rho) = R \sin(k\rho);\quad k^2 = \frac12\Lambda_6 + \frac14\beta^2
\label{uw1}
\eeq
as follows from solving the ($\mu\mu$) Einstein equation (\ref{mumu}), using 
(\ref{friedeq1}) to eliminate $\Lambda_4$.

More general solutions of type 3 cannot be found exactly, but an approximation is
possible for values of $E$ which are sufficiently close to the bottom of the potential in
fig.\ 1(a) so that it can be treated like a harmonic oscillator.  This gives
\beq
\label{soln4}
	M(\rho) \cong \left( z_0 + \epsilon\cos\left(k\rho\right)\right)^{2/5}
\eeq
where $z_0$ is the position of the minimum of the potential, and $k^2=U''(z_0)$.   In general $k$ and
$\epsilon$ are functions of $\Lambda_6$, $\Lambda_4$ and $\beta$. For example, in the case without
magnetic flux, $z_0 = ({2\Lambda_4/\Lambda_6})^{5/4}$ and $k^2=\frac12\Lambda_6$.  If one wants to
rescale the warp factor to unity, the physically meaningful place to do so is on the brane where the
observer is supposed to be, for example $z_0+\epsilon\to 1$. 

Interestingly, in the more general solution (\ref{soln4}), some dependence of the expansion rate on the
tensions $\tau_3$, $\tau_3'$ of the two 3-branes appears.  An application of the jump conditions to be
described in section 3.1 gives the following constraint:
\beq
	{\tau_3-2\pi \over \tau_3'-2\pi } = \left({z_0-\epsilon\over z_0+\epsilon}\right)^{3/2}
\eeq
This can be regarded as a fine tuning between the brane tensions required to obtain a given value
of the expansion rate determined by $\Lambda_4$.  There is a one-parameter family of
brane tensions which preserve the necessary relationship, showing that the expansion rate is
not uniquely determined by the brane tensions.  However the degeneracy of solutions along this
curve in the $\tau_3$-$\tau_3'$ plane does not constitute a solution to the cosmological constant
problem, and in this respect we disagree with the interpretation of \cite{Wett}.  In that reference,
it was assumed that a deficit angle could appear spontaneously at the position of one of the branes,
without having to specify any brane tension in the input to the construction of the model, namely
the stress-energy tensor.  Our point of view is that the 3-brane is not created by the geometry,
but rather any singularity in the 2D curvature is due to the presence of a 3-brane with nonvanishing
tension.  (Another difference between our work and that of \cite{Wett} is that we do not assume 
the deficit angle, hence brane tension, to be zero at $\rho=0$.  This enlarges the class of solutions we are considering.)

\section{Brane tensions and jump conditions}

In the previous section we generalized the analysis of \cite{RuSh} to include extra sources of
stress energy and to consider all possible signs of these various sources. A further new
ingredient which \cite{RuSh} did not consider is the presence of branes which bound the bulk
solutions given above.  In each case we wish to allow for a 3-brane at the origin of the two
extra dimensions, $\rho=0$, which could provide support for the standard model and our
observable 4D spacetime.  (A special case is a 3-brane with vanishing tension, which is
the same as no brane as far as Einstein's equations are concerned.)  The space will either
terminate with another 3-brane or a 4-brane at some maximum value $\rho_m$, or it may
be noncompact, according to the possiblities mentioned in section 2, although the latter
situation is not compatible with recovering 4D gravity at large distances.

\subsection{Jump condition at $\rho=0$}
In all cases, the constant of integration $R$ in (\ref{Lrel}) is related to the tension
$\tau_3$ of the 3-brane at $\rho=0$.  If $\tau_3$ is nonzero, there is a conical defect with
deficit angle $\delta$ given by $\delta = \kappa_6^2 \tau_3$, where $\kappa_6^2= 8\pi G_6$, in
terms of the 6D Newton's constant.  We can compute the ratio of the
circumference and the radius of an infinitesimal circle around the origin as 
$\lim_{\epsilon\to 0} 2\pi L(\epsilon)/\epsilon = 2\pi L'(0)$. Therefore, continuing to work
in units where $\kappa_6=1$, we have
\beq
	\tau_3 = {2\pi}(1-L'(0))
\eeq
If the space is terminated by another 3-brane at $\rho_m$ with tension $\bar\tau_3$,
the relation becomes
\beq
	\bar\tau_3= {2\pi}(1+L'(\rho_m))
\eeq

\subsection{Metric jump conditions at $\rho=\rho_m$}
On the other hand, if there is a 4-brane at $\rho_m$, it may need to be accompanied by
some additional form of stress energy in order to cut the bulk solution at this particular
location, because there are separate jump conditions for the $\theta\theta$ and $\mu\mu$
components of the Einstein equations.  Therefore one generally needs two tunable parameters to
satisfy both conditions.  These  conditions can be inferred from looking at the terms in the
Einstein equations which have second derivatives or delta functions:
\beqa
	\mu\mu&:&\quad {M''\over M} + \frac13{L''\over L} + {M'L'\over ML} + {M'^2\over M^2} = 
	-\frac13 
	T_0^{\ 0}\delta(\rho-\rho_m) + \dots  \\
	\theta\theta&:&\quad 2{M'^2\over M^2} + {4M''\over 3M} = -\frac13 
	T_\theta^{\ \theta}\delta(\rho-\rho_m) + \dots
\eeqa
We will impose $Z_2$ symmetry at the 4-brane so that the discontinuity in $M'$ is $-2M'$.  
Integrating in the vicinity of the delta functions, we find 
\beqa 
\label{jumpconds} 
	T_0^{\ 0} &=& 6{M'\over M} + 2{L'\over L}; \quad
	T_\theta^{\ \theta}  = 8{M'\over M}
\eeqa
where the functions are all evaluated at $\rho_m$. If the stress energy tensor was
given by a brane with pure tension, we would have $T_0^{\ 0} =  T_\theta^{\ \theta}$.  This
does not occur in the static AdS soliton solution (\ref{adssol}), where $M'/M - L'/L$ is
always nonzero, except as $\rho\to\infty$. One way to obtain a difference between $T_0^{\
0}$ and $T_\theta^{\ \theta}$ is to ``smear'' a 3-brane along the compact dimension of the
4-brane \cite{LMW}.  Since the 3-brane has $T_\theta^{\ \theta}=0$, this procedure will give an extra
contribution only to the $T_\mu^{\ \mu}$ components.  The shift in $T_0^{\ 0}$ due to
the smeared 3-brane tension $\tau_3'$ is given by 
  $T_0^{\ 0}-T_\theta^{\ \theta} =  {\tau_3'/[2\pi L(\rho_m)]}$.
This is not the only way to obtain a deviation $T_0^{\ 0}-T_\theta^{\ \theta}$; for example the Casimir energy
of a massless field confined to the 4-brane gives $T_0^{\ 0}-T_\theta^{\ \theta}\sim L^{-5}$ \cite{CN}.
In general, the energy density of the source $\tau'$ can scale like $1/L^\alpha(\rho_{m})$, where $\alpha$ determines its equation of state.  Conservation of stress-energy then
dictates its contributions to the different components of the stress tensor \cite{Luty},\cite{BCCF}:
\beqa
	T_0^{\ 0} &=& T_4 +V_m(\phi) + {\tau'\over L^\alpha(\rho_m)}\\
	T_\theta^{\ \theta} &=& T_4 + V_m(\phi) + (1-\alpha){\tau'\over L^\alpha(\rho_m)}
\eeqa
The scalar field potential $V_m(\phi)$ is explained in the next subsection.

\subsection{Jump conditions for $\phi$}
In the absence of delta function sources for $\phi$, regularity of the solutions requires
that $\phi'(0) = \phi'(\rho_m)=0$.  If potentials $V_0(\phi)\delta^{(2)}(\rho),\ V_m(\phi)
\delta(\rho-\rho_m)$ are included in the Lagrangian, $\phi'$ may be nonzero at these points.
The effect of $V_m$ is easily found by integrating the $\phi$ equation of motion in the 
vicinity of $\rho_m$ assuming $Z_2$ symmetry:
\beq
\label{phijc}
	\phi'(\rho_m) = -\frac12 {dV_m\over d\phi}
\eeq
The effect of $V_0$ is more subtle.  One can analyze it by regularizing the 2D delta function,
$\delta^{(2)}(\rho)\to 1/(\pi L'(0)\epsilon^2)$ for $\rho<\epsilon$, and solving the field
equation in the small $\rho$ region using regular boundary conditions, $\phi'(0)=0$.  The
behavior of $\phi$ depends on the choice of potential.  For $V_0 = \lambda(\phi-\phi_0)^2$,
we find that $\phi\sim  c_1 J_0(m \rho) + c_2 Y_0(m \rho)$ in the interior region, with
$c_2 \sim 1/\ln(\epsilon)$.  Therefore in the limit $\epsilon\to 0$, we recover the boundary
condition $\phi'(0)=0$ despite the presence of the potential.  On the other hand, the linear
potential $V_0 = \lambda \phi$ yields a singular solution, $\phi \sim \phi_0 \ln(\rho)$ near
the origin.

\section{Stabilization in warped model} 

Our ultimate goal is to understand the rate of expansion of the universe in models where we
live on a codimension-two brane, with the expectation that the Hubble rate is due to the energy
density of this 3-brane.  Since the present universe has an energy density which is much less
than that corresponding to the Planck scale, it is appropriate to treat $\Lambda_4$ as a
perturbation.  We therefore want to first understand the model in the static limit,
$\Lambda_4=0$.  Studies of 5D models have underscored the point that it is essential to
stabilize massless volume moduli in order to recover 4D Einstein gravity at low energies, so we
now consider a stabilization mechanism, in this section for the case where the 6D spacetime
is warped by a negative bulk energy density.  

\subsection{Stabilization by bulk scalar}
It has been shown \cite{BCCF} that the radial size of the extra dimension is unstable 
in the AdS soliton model; one way to stabilize it is with a bulk scalar field which is 
prevented from vanishing by the
presence of a potential $V_m(\phi)$ on the 4-brane at $\rho_m$.  Following \cite{CFNW}, we
make the simple choice $V_m(\phi) = -\lambda\phi$, which gives the boundary condition
$\phi'=\lambda/2$, as noted in (\ref{phijc}).  In the large $\rho$ region, where $M\sim L\sim
e^{2k\rho/5}$.  In the large and small $\rho$ regions, the solution for $\phi$ is approximately
\beqa
	\phi &=& B_1 e^{\sigma_+\rho}+ B_2 e^{\sigma_-\rho},\qquad 
	\phantom{A}\rho\gg 1/k\nonumber\\
	\phi &=& \phi_0(1 + m^2\rho^2/4 + \dots),\quad \rho \ll 1/k
\eeqa
with $\sigma_\pm = -k\pm\sqrt{k^2+m^2}$ and $k = \sqrt{-5\Lambda_6/8}$.  The behavior at
$\rho=0$ follows from assuming there is no potential for $\phi$ at $\rho=0$. By matching the
two solutions at $\rho\sim 1/k$, and further assuming that $m\ll k$, one finds that $B_2 = 
O(m^4/k^4)\phi_0$, $B_1\cong\phi_0$.  Ignoring the $B_2$ term, 
the jump condition at $\rho_m$ fixes 
$\sigma_+\phi(\rho_m) = \lambda/2$.  The
assumption $m\ll k$ is useful for limiting the back reaction of the scalar on the geometry; in
the limit $m\to 0$, $\phi'\to 0$, and the effect of the scalar is to shift the value of
$\Lambda_6$, so that the solution for the metric would have the same form as that without the
scalar.  In this regime of parameters, the scalar solution can be approximated as $\phi\cong
\phi_0e^{m^2\rho/2k}$.

In solving the jump conditions for the metric, it is useful to consider linear combinations
which are independent of either the extra source $\tau'$ or the bulk vacuum energy $\Lambda_6$.
These linear combinations are
\beqa
\label{jc1}
	-\lambda\phi + T_4 &=& \left(6+{2\over\alpha}\right)\,{M'\over M}
	 + \left(2-{2\over \alpha} \right) \,{L'\over L} 
	\equiv \left(6+{2\over\alpha}\right) \m
	 + \left(2-{2\over \alpha} \right) \l
	\\
\label{jc2}
	{\alpha\tau' \over 2L^\alpha} &=& {L'\over L} - {M'\over M} 
	\phantom{AAAAAAAAAAAa}\equiv \l-\m
\eeqa
where all quantities are to be evaluated at $\rho=\rho_m$.  The right hand side depends only
on ratios where the overall normalization of the metric elements cancel out, whereas the
left hand side depends on $L$ itself.  Using $L=R M'$, $L'=R M''$,
$L'(0) = 1-\tau_3/2\pi$ and (\ref{thth}, \ref{adssol}) to evaluate $M''(0)=2k^2/5$, we can show that
the constant $R$  equals $(5/2k^2)(1-{\tau_3\over 2\pi})$.  (Recall that we have
rescaled the 4D coordinates $x_\mu$ so that $M(0)=1$).  
Then $L$ can be expressed as
\beq
\label{LMrel}
	L = \frac52 k^{-2}\left(1-{\tau_3\over 2\pi}\right) M' \cong 
	k^{-1}\left(1-{\tau_3\over 2\pi}\right) M
\eeq
where the latter approximation holds for $\rho\gg 1/k$.
Eqs.\ (\ref{jc1}, \ref{jc2}) determine $\rho_m$, the position of the 4-brane, and they give a 
constraint on some function of $T_4$, $\tau_3$ and $\tau'$.  The latter is the fine-tuning
which must be done in order to obtain a static solution.  We will carry this out explicitly
below.

\subsection{Back reaction of scalar (or magnetic flux) on metric}
To solve the jump conditions (\ref{jc1}, \ref{jc2}), we need to know how the scalar perturbs
the metric.  Although we will consider stabilization by the scalar and not the flux, for 
generality we indicate how the same analysis can be carried out for stabilization by flux.
Similarly, we can also compute the perturbation away from the static solution by keeping the
effect of the 4D cosmological constant, even though our immediate goal is to examine the
static solution.

Let $\delta\m$ be the perturbation to $M'/M$ in eq.\ (\ref{thth}), which 
when linearized takes the form 
\beq
	\delta\m' + 5\m\,\delta\m = \frac18\left(4{\Lambda_4 \over M^{2}} 
	- \phi'^2-m^2\phi^2 + {n^2\phi^2\over L^2}+{\beta^2\over M^{8}}\right)
\eeq
The part of $\delta\m$ in the absence of flux, winding, or cosmological expansion, due only
to the stabilization mechanism, is
\beqa
	\delta\m_\phi &\equiv& \delta\left({M'\over M}\right) 
= -\frac18 M^{-5}\int_0^\rho d\rho\, (\phi'^2+m^2\phi^2) M^5\nonumber\\
\label{msol}
	&\cong& -{m^2\phi_0^2\over 64\cosh^2(k\rho)}\left( {e^{2(k+\sigma_+)\rho} - 1\over k+\sigma_+}
	 + 2 {e^{2\sigma_+\rho} - 1\over \sigma_+} + {e^{2(-k+\sigma_+)\rho} - 1\over -k+\sigma_+}
	\right)
\eeqa	
It can be shown that in the $m\ll k$ limit in which we are interested, it is always true
that $\phi'^2\ll m^2\phi^2$, so we dropped the
derivative term to obtain  (\ref{mlsol}). Similarly the decaying exponential part of $\phi$ gives only subdominant contributions
in $m^2$, so we approximated $\phi = \phi_0e^{\sigma_+\rho}$.  Now
eq.\ (\ref{msol}) can be integrated again to find the deviation in $M$ itself,
\beqa
\label{deltaMphi}
	{\delta M_\phi\over M} &=& \int_0^\rho d\rho\, \delta\m_\phi 
	\cong -{\lambda^2\over 16m^2} k\rho_m \ (\hbox{at\ } \rho=\rho_m)
\eeqa
where in the last approximation we have kept only the leading behavior in $m^2$.
The last result shows that $\lambda^2 k\rho_m/m^2$ must be kept small in order for the
perturbation to be under control.
Notice that we have used our freedom to impose the boundary condition $\delta M_\phi(0) = 0$ 
to preserve the normalization $M(0)=1$.

To compute the analogous shift $\delta\l_\phi$ in $L'/L$, we perturb eq.\ (\ref{rr}).  In 
fact it will be convenient to compute $\delta\l_\phi - \delta\m_\phi$,
\beq
\label{lmsol}
	\delta\l_\phi - \delta\m_\phi = -\left(4+{\l\over\m}\right) \delta\m_\phi + 
	{1\over 8\mu}(\phi'^2-m^2\phi^2)
\eeq
Using $\l/\m = (5\coth^2(k\rho) - 3)/2$ and $\mu = (2k/5)\tanh(k\rho)$ for the unperturbed
quantities, we can explicitly integrate (\ref{lmsol}) from 0 to $\rho$.  The result is
\beq
\label{LMsol}
	{\delta L_\phi\over L}  - {\delta M_\phi\over M} = \frac5{16} {\lambda^2\over m^2}
	\left( e^{-2\sigma_+\rho} - {1\over 2 + m^2/k^2} + O(e^{-2k\rho}) \right)
\eeq
This result will be essential for finding the size of the radial extra dimension
$\rho_m$ from the stabilization mechanism.

\subsection{Solution of jump conditions}
We are now ready to solve for $\rho_m$ and the fine-tuning between stress-energy components
which leads to the static solution.  Toward this end, it is useful to rewrite the difference
between eqs.\ (\ref{thth}) and (\ref{mumu}) in the form
\beq
\label{mleq}
	\left(M^4 L (\m-\l)\right)' = M^4L\left({\Lambda_4\over M^2} 
	+ {n^2\phi^2\over L^2} + {\beta^2\over M^8}	\right)
\eeq
If we are only interested in stabilization by the Goldberger-Wise mechanism, and the static
solution, then the right hand side of (\ref{mleq}) vanishes, leading to
\beq
\label{mlsol}
	\l-\m = {c\over M^4 L} =  {L'(0)\over M^4 L} 
\eeq
where we evaluated the equation at $\rho=0$ to evaluate the constant of integration $c$.
Combining (\ref{mlsol}) with the jump condition (\ref{jc2}), we find the equation which
determines the $\rho_m$, namely $2L'(0)/\alpha\tau' = M^4 L^{1-\alpha}$.   
The only hope for obtaining the desired large hierarchy $\rho_m \gg 1/k$ is in
the case where $\alpha=5$.  Otherwise we obtain $e^{2(1-\alpha/5)k\rho_m} \sim L'(0)/\tau'$,
which gives a large hierarchy only by tuning $\tau'$ to be unnaturally small in units of
the 6D gravity scale.  On the other hand, in the case $\alpha=5$, we must rely on the
small correction (\ref{LMsol}) to find the leading $\rho$ dependence in the jump condition,
\beq
\label{rhoeq}
	y\equiv {2\over 5\tau'}\left(1-{\tau_3\over2\pi}\right) = \left({M\over L}\right)^4
	= 1 - 4\left({\delta L_\phi\over L}  - {\delta M_\phi\over M} \right) 
	+ O(e^{-2k\rho_m}) 
\eeq
The result for $\rho_m$ is
\beq
	\rho_m = -{k\over m^2}\ln\left({4m^2\over 5\lambda^2}(1-y) + \frac12\right)
\eeq
This solution only exists if $\tau'$ lies in the range
\beq
\label{tauprange}
   {2\left(1-{\tau_3\over2\pi}\right)\over 5\left(1 + {5\lambda^2\over 8 m^2}\right)}
  < \tau' < {2\left(1-{\tau_3\over2\pi}\right)\over 5\left(1 - {5\lambda^2\over 8 m^2}\right)}
\eeq
This range could have had a large upper limit if ${5\lambda^2/8 m^2}\lsim 1$.  However we found
in (\ref{deltaMphi}) that in fact we must have $\lambda^2/m^2 < 1/(k\rho_m)$ in order for
perturbation theory to not break down.  So we see that there is still a fine tuning on $\tau'$
for this model to give a large hierarchy.  It is qualitatively different from the kind of
tuning that would occur in models with $\alpha<5$; for these one would need to make the
magnitude of $\tau'$ exponentially smaller than its natural scale in terms of the 6D gravity
scale.  For $\alpha=5$, the magnitude of $\tau'$ can be natural, but it must be very close to a
specific value.  One virtue of the scalar field is that it reduces the amount of tuning which
is necessary.   A similar tuning was present in the absence of the scalar field: (\ref{rhoeq})
shows that $y$ must be exponentially close to $1$ if $\phi$ is absent, and if we insist on
getting a large hierarchy. Once $\phi$ is introduced, $y$ must still be close to 1, but no
longer with exponential precision.

We could simply give up the tuning of $\tau'$ and live with a model which does not try to
solve the hierarchy problem.  But since the main purpose of this paper is to explore the
strange cosmological properties of this kind of model, questions of naturalness of the choice
of parameters are secondary.  It is convenient to adhere to the strongly warped case, since
the hierarchy simplifies the algebraic analysis.  We have seen that even in the completely
unwarped case, similar behavior is revealed, concerning the unusual dependence of the Hubble
rate on the brane tension is concerned.

The remaining jump condition (\ref{jc1}) is the fine tuning that must be imposed to make the
effective 4D cosmological contant $\Lambda_4$ vanish:
\beq
	T_4 -{\lambda^2 k^2\over 2 m^2} - {16\over 5}k = 5\tau' e^{-2k\rho_m}\left(
	1+ {5\lambda^2\over 16 m^2} k\rho_m \right)
\eeq
underscoring the fact that ultimately one needs an additional mechanism to fully solve the
cosmological constant problem in these models. {Our purpose here is 
not to address the problem of the 4-D cosmological constant, but rather the 
possible cosmological solutions of codimension-2 braneworlds.}

\subsection{Stabilization by magnetic flux?}
We can attempt to stabilize the warped solution using magnetic flux; however since the
effects of the flux are highly suppressed at large $\rho$ because of the large exponential
factor $1/M^8$, it is not clear that it actually can stabilize the highly warped solution.  Following
the same steps as for the scalar field, we find 
\beqa
	\delta\mu_\beta &=& {\beta^2\over 8\cosh^2(k\rho)}\int_0^\rho d\rho M^{-3} 
	\cong   {\beta^2\over 8k\cosh^2(k\rho)}\left(c_0 + c_1e^{-6k\rho/5} + c_2 e^{-c_3
	k\rho}\right)\nonumber
\eeqa
where the approximate fit has $c_0=1.38$, $c_1=-1.91$, $c_2 = -c_0-c_1$, $c_3=-(1+\frac65
c_1)/c_2$ and in the large $\rho$ limit
\beqa
	{\delta L_\beta\over L}  - {\delta M_\beta\over M} \cong \frac{5\beta^2}{16k^2} \left(
	d_0 + d_1 e^{-2k\rho}\right)\nonumber
\eeqa
where $d_0 =-0.66$, $d_1=5.5$.  The resulting equation for $\rho_m$ has the same form as
(\ref{rhoeq}), except now the term ${\delta L_\beta/L}  - {\delta M_\beta/M}$ depends on
$\rho_m$ only through $e^{-2k\rho_m}$, which is of the same order as terms we ignored in
(\ref{rhoeq}).  As explained there, this kind of dependence does not allow for a large
hierarchy unless $\tau'$ is tuned with exponential precision to a special value.  This is
exaclty the same  situation as in the model with no stabilization, leading one to suspect that 
magnetic flux does not actually stabilize this model, unless the warping is weak
(which would be the case if $\Lambda_6 >0$).

\section{Friedmann equation in warped model}
We now come to the main point of the paper, in the context of the AdS-type models: the rate of expansion in the models we have
described depends on the energy density (tension) of the standard model 3-brane in a strange
way, relative to our expectations from 4D general relativity.  Such an outcome would be less
surprising if the low energy limit of the model was a scalar-tensor theory of gravity, in which
the coupling of the extra massless scalar to the brane tension could explain the exotic
effect.  However we will note that the conclusion is valid even when moduli have been
stabilized.  We now demonstrate this in the warped case of the expanding  AdS soliton
model, (\ref{adssol2}), supplemented by a scalar field which stabilizes the radion by the
Goldberger-Wise mechanism \cite{GW,CN,CFNW,BCCF}.

The main work has already been done in the previous section; there we found the back reaction
to the metric by treating the scalar field as a perturbation.  In the same way we can find the
perturbation to the static solution due to a small rate of expansion by treating the
$\Lambda_4/M^2$ term to first order.  We have a double expansion, in $\Lambda_4$ and in 
$\phi^2$.  We will work to first order in either quantity, but not including the mixed terms of
order $\Lambda_4\phi^2$.

The perturbation $\delta\mu_{\sss\Lambda}$ analogous to
(\ref{msol}) is 
\beqa
\label{msol2}
	\delta\m_{\sss\Lambda} &=& \frac{1}{2} M^{-5}\int_0^\rho d\rho\, 
	{\Lambda_4\over M^2} M^5 \\
\label{mlsol2app}
		  &\cong&  {5\Lambda_4\over 12\cdot 2^{6/5} k}\cosh^{-2}(k\rho) 
	\left( e^{6k\rho/5}-1 - a(e^{-kb\rho}-1)\right)
\eeqa
where $a = (2^{6/5}-1)^2$, $b = (6/5)(2^{6/5}-1)^{-1}$.
The integral cannot be done analytically, but the appproximation (\ref{mlsol2app}) is chosen
to give the correct large $\rho$ behavior, and it matches the exact value and its first
derivative at $\rho=0$; it is accurate to $0.6\%
$ everywhere, as shown in fig.\ 3.

\FIGURE{
\centerline{\epsfxsize=5.in\epsfbox{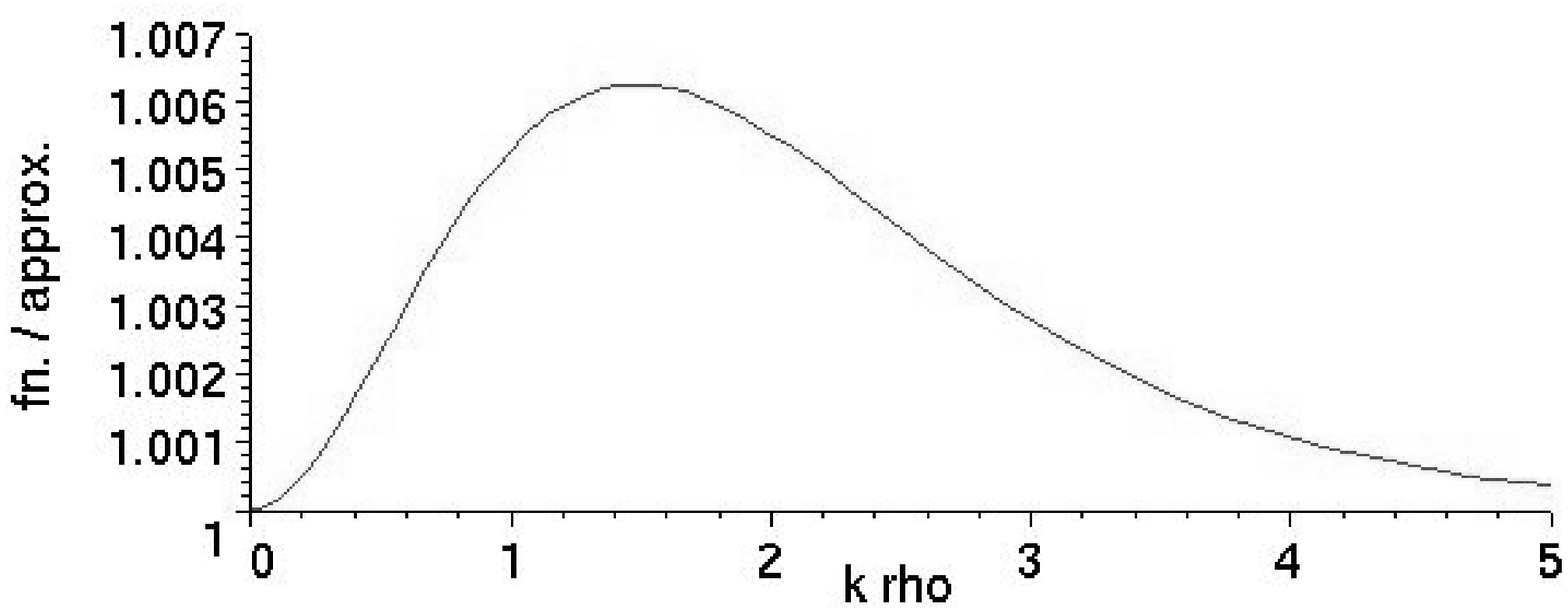}}
\caption{\small
Ratio of the exact value of the integral in (\ref{mlsol}) to the approximation
(\ref{mlsol2app}) as a function of $k\rho$.}
}

The shift in $\delta\l$ is given by
\beq
\label{lmsol2}
	\delta\l_{\sss\Lambda} - \delta\m_{\sss\Lambda} = 
	-\left(4+{\l\over\m}\right) \delta\m_{\sss\Lambda}  + 
	{\Lambda_4\over 2\mu M^2}
\eeq
By numerical evaluation and empirically fitting, we find that the following expression is
a very good approximation to the integral of (\ref{lmsol2}):
\beq
\label{LMsol2}
	{\delta L_{\sss\Lambda}\over L}  - {\delta M_{\sss\Lambda}\over M} \cong \frac54 {\Lambda_4\over k^2}
	\left(-1 + \cosh^{-\frac45}(k\rho) \right)
\eeq
This captures with great accuracy asymptotic approach of the solution to its large $\rho$
limit, as well as having the correct small $\rho$ behavior. The comparison between the exact
result and the approximation is shown in fig.\ 4.

\FIGURE{
\centerline{\epsfxsize=3.5in\epsfbox{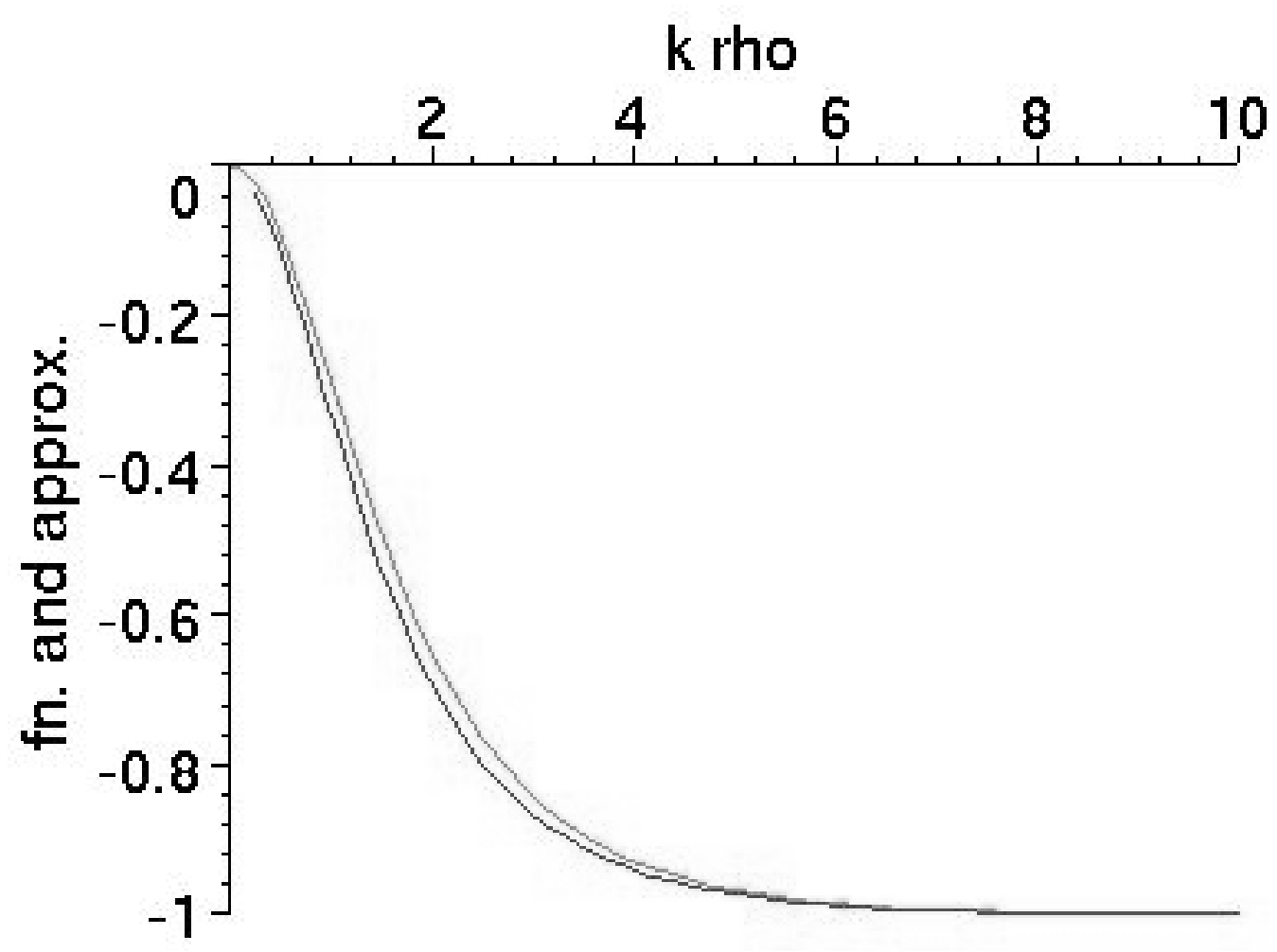}}
\caption{\small
Numerical integration of (\ref{lmsol2}) versus $k\rho$, and the approximation
(\ref{LMsol2}).}
} 	

The new contribution to the jump condition determining the combination $\l-\m$ from (\ref{mleq})
gives
\beqa
	\l-\mu &=& {1\over M^4 L}\left( L'(0) -  \int_0^\rho d\rho\, {\Lambda_4\over M^2} M^4 L 
	\right)\nonumber\\
	&=& {L'(0)\over M^4 L}\left( 1 - {5\Lambda_4\over 6 k^2}(M^3-1)\right)
\eeqa	
where we used (\ref{LMrel}) to perform the integral exactly.  This modifies (\ref{rhoeq})
by giving a new term proportional to $\Lambda_4$:
\beq 
\label{yeq}
	y\equiv {2\over 5\tau'}\left(1-{\tau_3\over2\pi}\right) \cong
	1 + {5\lambda^2\over 16}\left(e^{-2\sigma_+\rho}-\frac12\right) + 
	{5\Lambda_4\over 6 k^2} M^3 
\eeq
where we have ignored all exponentially small corrections.  

Let us now recall what is the origin of the expansion of the universe according to our
assumptions.  We are interested in starting from a static configuration, and then perturbing it
by adding some energy density on the 3-brane.  Of course it is possible that the expansion of
our own universe could be due to a small mismatch of bulk energy densities which are
distributed in the extra dimensions.  But our specific interest in this paper is to see how 
the energy density on the 3-brane affects the Hubble expansion.  Therefore we will make the
assumption that the quantity whose value is relaxed relative to the static fine-tuned situation
is $\tau_3$; hence in our perturbation series $\delta\tau_3$ and $\Lambda_4$ are of the same
order.  When comparing (\ref{yeq}) to the corresponding relation (\ref{rhoeq}) for the static
solution, there is one other quantity which must in general vary: the position of the 4-brane
$\rho_m$ will deviate by some amount $\delta\rho_m$.  However, all terms containing
$\delta\rho_m$ are or order $\Lambda_4\phi^2$ or $\Lambda_4^2$ and so can be neglected.
Therefore the difference between  (\ref{yeq})
and (\ref{rhoeq}) gives
\beq
\label{friedeq2}
	\framebox{$\displaystyle	\Lambda_4 = -{3k^2\over 5 \pi}\, {\delta\tau_3\over
	1-{\tau_3\over 2\pi}} \cosh^{-\frac65}(k\rho_m)$}
\eeq
where we used the constraint (\ref{tauprange}) to eliminate $\tau'$.
This is the result we have been seeking, which shows an unexpected relation between $\Lambda_4$
and $\delta\tau_3$: a {\it decrease} in the energy density on the 3-brane leads to cosmological
expansion in this model.  

It is straightforward to give the more general version of eq.\ (\ref{friedeq2}), for 
arbitrary values of $\alpha$, and without assuming that we are perturbing around a 
static solution.  We find that
\beq
	\Lambda_4 = {6k^2\over 5(M^3-1)}\left( 1-  {\alpha M^4 L^{1-\alpha}\tau'
	\over 2 \left(1-{\tau_3\over 2\pi}\right) }\right)
\eeq
where $M$ and $L$ are to be evaluated at $\rho_m$.  This shows that the surprising dependence
on $\tau_3$ is not due to fine tuning of parameters.  The Hubble rate is a decreasing function
of $\tau_3$ quite generally in this model.  By choosing $\tau_3$ so as to make $\Lambda_4=0$,
and perturbing $\tau_3\to \tau_3+\delta\tau_3$, one recovers eq.\ (\ref{friedeq2}).

When we perturb the other jump condition (\ref{jc1}), we obtain an equation for the shift
in $\rho_m$,
\beq
	\left(\m'+ \frac15(\l'-\m')\right) \delta\rho_m + \delta\m_{\sss\Lambda} + 
	\frac15(\delta\l _{\sss\Lambda} - \delta\m_{\sss\Lambda}) = 0
\eeq
which gives
\beq
	\delta\rho_m = {-5\Lambda_4\over 32k^3}\, {e^{2k\rho_m}\over\cosh^{\frac45}(k\rho_m)}
	= {3\over 20\pi}\, {\delta\tau_3\over k\tau'}
\eeq

\section{Stability of unwarped model}
As already pointed out in eq.\ (\ref{uw1}), it is possible to find unwarped solutions  \cite{CG,nav} 
where the bulk is a two-sphere stabilized by magnetic flux. In these models the expansion 
rate is insensitive to the three brane tension.  To be sure that this unusual behavior is not related
to the presence of unstabilized moduli, in this section we perform a more complete analysis of the
stability of the model than was done in ref.\ \cite{CGHW}.
In the following we will set $M=1$. We will consider in detail the static case and then 
give the main results of the analysis in the presence of the expansion. 

In the static case $\Lambda_{4} =0$ and, from eqs.\ (\ref{mumu})--(\ref{thth}) we have
\beq 
\ell^2 +\ell' = -\beta^2 ,~~~~~~~~~~~~ \Lambda_{6} = \frac{\beta^2}{2}.
\label{uw2} 
\eeq
The fluctuations of this model may be classified according to four-dimensional 
Lorentz transformations. On top of the usual transverse and traceless 
tensor  (i.e. $\partial_{\mu} h^{\mu}_{\nu} = h_{\mu}^{\mu} =0$) corresponding 
to five degrees of freedom there are three divergenceless vector modes (corresponding to nine
degrees of freedom) and seven scalar modes. Overall, the perturbed 
six-dimensional metric has 21 degrees of freedom. These 21 fluctuations of different spin 
transform under coordinate transformations parametrized by the infinitesimal shift 
$\epsilon_{A} \equiv (\epsilon_{\mu}, \epsilon_{\rho},\epsilon_{\theta})$.
The infinitesimal shift $\epsilon_{\mu}$ along the four-dimensional spacetime 
can be decomposed, in turn, as the derivative of a scalar and a divergenceless 
vector, i.e. 
$\epsilon_{\mu} = \partial_{\mu}\epsilon + \zeta_{\mu}$. It is now easy to see that 
there are 3 scalar gauge functions ($\epsilon_{\rho}$, $\epsilon_{\theta}$ and $\epsilon$)
and one vector gauge-function. Of the seven scalar degrees of freedom, 
three of them can be gauged away by using the scalar gauge functions. Alternatively,
the seven gauge-dependent scalars can be rearranged 
into four gauge-invariant scalar fluctuations. In the context of the study of six-dimensional 
Abelian vortices \cite{MG1,MG2} it was shown that a convenient gauge choice brings the 
line element in the form 
\beq
ds^2 = (1 - 2\psi) \eta_{\mu\nu} dx^{\mu} dx^{\nu} + ( 1 + 2 \xi) d\rho^2 + L(\rho)^2 ( 1 -1 2\varphi) d\theta^2
+ 2 \pi L(\rho) d\rho\, d\theta 
\label{lineel}
\eeq
where $\eta_{\mu\nu}$ is the Minkowski metric. Since all the scalar gauge functions are 
completely fixed, no spurious gauge modes will appear. Furthermore, it can be shown that 
the scalar fluctuations appearing in (\ref{lineel}) obey the same evolution equations 
of the four gauge-invariant fluctuations which have been defined, in general terms, in \cite{MG2}.

The four scalar fluctuations of the geometry are coupled, through the perturbed Einstein equations
\beq 
\delta^{} R_{AB} = \delta^{} \tau_{AB},~~~~~~~~~~~~~ \tau_{AB} = T_{A B} - \frac{1}{4} T_{C}^{\ C} G_{A B},
\label{perteineq}
\eeq
to the scalar fluctuations of the sources whose 
evolution can be obtained by perturbing to first order 
\beq 
\partial_{A} \biggl( \sqrt{|G|} F^{A B}\biggr) =0.
\label{pertmaxeq}
\eeq
The two relevant fluctuations of the source are the fluctuations 
of the $\rho$ and $\theta$ components of the vector potential. These fluctuations will be donoted as 
$A_{\rho}$ and $A_{\theta}$. The divergence-full part of the four-dimensional vector potential 
is also a scalar but it can be gauged away by using the 
$U(1)$ gauge symmetry.

Now that the gauge is fully fixed, we are ready to find the spectrum of
fluctuations for the class of unwarped models defined by (\ref{uw2}).  In connection with the
automatic adjustment of the solutions to cancel out any dependence on the tension of the
3-branes, we are specifically interested in the radial modes and, therefore, the possible dependence 
of the perturbed quantities upon the coordinate $\theta$ will be ignored.
Using (\ref{lineel}) the various components of eq.\ (\ref{perteineq}) are  
\beqa
\label{munuper}
\negsp{\mu\neq\nu:}\quad &&\xi = \varphi + 2 \psi,\\
\label{mumuper}
\negsp{\mu=\nu:}\quad && \psi'' +\ell \psi' + \Box \psi = -\frac{\beta}{2 L} A_{\theta}' + \frac{\beta^2}{2} (\xi - \varphi),\\
\label{rrper}
\negsp{\rho\rho:}\quad && \varphi'' + 4 \psi'' + \ell ( 2 \varphi' + \xi') -\frac{3}{2} \beta^2 \varphi - \Lambda_{6} \xi - \Box\xi = 
\frac{3}{2} \frac{\beta}{L} A_{\theta}' \\
\label{ththper}
\negsp{\theta\theta:}\quad && \varphi''  +\ell ( 2 \varphi' + \xi' + 4 \psi') - \frac{3}{2}\beta^2 \varphi + \Box\varphi + 
2 \biggl[ \ell^2 + \ell' + \frac{3}{4} \beta^2 \biggr] \xi = \frac{3}{2} \frac{\beta}{L} A_{\theta}',\\
\label{murper}
\negsp{\mu\rho:}\quad && \varphi' + \ell (\varphi +  \xi) + 3 \psi' = \frac{\beta}{L} A_{\theta} ,\\
\label{muthper}
\negsp{\mu\theta:}\quad && \pi' + 2 \ell \pi  = -\frac{\beta}{L} A_{\rho},\\
\label{thrper}
\negsp{\theta\rho:}\quad && \Box \pi =0.
\eeqa
where $\Box$ denotes the four-dimensional D'Alembertian, $-\partial_t^2 + \partial_{\vec x}^2$.
Eqs.\ (\ref{munuper})--(\ref{thrper}) should be supplemented by the evolution equations 
describing the fluctuations of the sources which can be obtained by perturbing 
eq.\ (\ref{pertmaxeq}) to first order:
\beqa
&& A_{\theta}'' - \ell A_{\theta}' + \beta L ( 4 \psi' + \xi' - \varphi') + \Box A_{\theta} =0,
\label{ath}\\
&&\Box A_{\rho} =0, \qquad A_{\rho}' =0.
\label{Arho}
\eeqa

Notice that $\pi$ and $A_{\rho}$ are decoupled from the other equations.  The resulting system of
coupled equations (\ref{thrper}) and (\ref{Arho})  implies that both $\pi$ and $A_{\rho}$ are massless
excitations. Furthermore, since  $A_{\rho}$ is constant, eq.\ (\ref{muthper}) can be easily solved with
the result that  
\beq  
\label{pieq}
	\pi(\rho) = \frac{c_{\pi}}{L^2} - \beta A_{\rho} \int d\rho L(\rho). 
\eeq
However this mode is not normalizable, hence it is unphysical. In the action perturbed to second order
the kinetic term for $\pi$ appears multiplied by only one factor of  $L$. Thus the kinetic term for
$\pi$ diverges as $\rho\to 0$,  in agreement with the findings of \cite{MG2}. 

The remaining longitudinal degrees of freedom (i.e. $\varphi$, $\psi$, $\xi$ and $A_{\theta}$) 
satisfy a coupled set of linear 
differential equations whose eigenvalues determine the stability 
of the solution (\ref{uw2}). The boundary conditions to be imposed at 
the origin are \cite{BCCF}:
\beq
\varphi(0) + \xi(0) =0, ~~~~~\psi'(0)=0,~~~~~~A_{\theta}'(0) =0.
\label{boundary}
\eeq
If we now consider eq.\ (\ref{munuper}) we see that, for $\rho\to 0$,
eqs.\ (\ref{boundary}) imply 
\beq 
\phi(0) + \psi(0) =0
\label{diff}
\eeq
It is convenient to define two new variables:
\beq 
\psi = \frac{X  - Y}{2}, ~~~~~~~~~~\xi = \frac{X + Y}{2}.
\label{XYdef}
\eeq
With these variables we will have that the linear combination 
$3 \times(\mu,\nu) + (\theta,\theta)$ leads to 
\beq 
X'' + 3 \ell X' + (\Box-\beta^2) X = \beta^2 Y 
\label{X}
\eeq
while the difference of the 
$(\rho,\rho)$ and $(\theta,\theta)$ components of the perturbed Einstein equations
(\ref{rrper}) and (\ref{ththper}) gives
\beq
X'' - Y''  - \ell (X' - Y') = \Box Y
\label{XY}
\eeq
and eq.\ (\ref{murper}) determines the gauge field
\beq 
A_{\theta} = \frac{L}{\beta} ( X' + 2 \ell Y).
\eeq
In terms of $X$ and $Y$ the boundary conditions (\ref{boundary}) are now $X'(0)=Y'(0)$ and 
$Y(0)=0$. 

It is easy to verify that one solution to the coupled equations (\ref{X}, \ref{XY}) matches the
results of ref.\ \cite{CGHW}, with $X=$ constant, $Y=0$, and mass eigenvalue
\beq
\label{radmass}
	m^2_r = \beta^2
\eeq
which is the radion mass squared of this model.  Although the analysis of \cite{CGHW} was in a
4D effective theory, their result is exact since the solution has no dependence on the extra
dimension.  The constancy of the wave function shows that this must indeed be the ground state
of the system (\ref{X}, \ref{XY}), since nonconstant solutions should have higher energy and
correspond to Kaluza-Klein excitations of the ground state.  Nevertheless to be thorough we have
conducted a numerical search for other eigenstates using the shooting method, and verified that
(\ref{radmass}) is indeed the ground state solution.
  
The same analysis can be done in the case where the background solution has
de Sitter expansion, as in eq.\ (\ref{uw1}), rather than being static.
The system (\ref{X}, \ref{XY}) becomes
\beqa
&& X'' +3 \ell X'+X\left(\Box-\beta^2+\frac{33}{4}H^2\right) = \left(\beta^2+12H^2\right)Y \\
&& X''-Y'' -\ell (X'-Y')= \left(\Box 
+\frac{9}{4}H^2\right)
\eeqa
The form of the bulk solution is the same as when $H=0$, but the mass eigenvalue generalizes to
 \beqa
m_r^2 = \beta^2 - \frac{33}{4}H^2,
\eeqa
showing that a sufficiently large rate of expansion destabilizes the compactification.

Let us now discuss the vector modes, which must be considered in  a fully consistent decomposition of
the metric and source fluctuations 6D. It was recently shown in \cite{nav} that a massless graviphoton
field is present in the low-energy spectrum.  In the following we restrict ourselves to the spectrum of
the vector excitations with only radial dependence. The $\theta$ dependence of the perturbed
quantities, considered in \cite{MG1}, would presumably correspond to KK excitations of the
lowest modes which we seek here.

Denoting by $V_{\mu}$ and $Z_{\mu}$ the divergenceless 
graviphoton fields appearing, respectively, in the 
$(\mu,\rho)$ and $(\mu,\theta)$ components of the perturbed metric, 
the equations for the coupled system of vector fluctuations reads:
\beqa
\label{munuperv}
\negsp{\mu\neq\nu:}\quad && V_{\mu}' + \ell V_{\mu}=0 ,\\
\label{murperv}
\negsp{\mu\rho:}\quad && \Box V_{\mu} =0,\\
\label{muthperv}
\negsp{\mu\theta:}\quad && Z_{\mu}'' + \ell Z_{\mu}' + \ell' Z_{\mu} + \Box Z_{\mu} + 2 \beta A_{\mu}'=0 \\
\label{muv}
\negsp{\mu:}\quad && A_{\mu}'' + \ell A_{\mu}'  + \Box A_{\mu} - \beta[ Z_{\mu}' + \ell Z_{\mu} ] =0,
\eeqa
where $A_{\mu}$ is the divergenceless fluctuation of the vector potential and eq.\ (\ref{muv}) follows 
from the perturbed component of the evolution equation of the gauge field.

From eq.\ (\ref{murperv}) it follows that $V_{\mu}$ is always massless in perturbation theory, as was
argued  in \cite{nav}. This conclusion, valid for radial excitations, follows from the cancellation  of
the contribution of $\Lambda_{6}$ and the magnetic flux in the perturbed equations. However eq.\
(\ref{munuperv}) implies that $V_{\mu} \sim L^{-1}$. Like the $\pi$ mode in (\ref{pieq}) this mode
is not normalizable and therefore unphysical.

To analyze the coupled system of excitations for $Z_{\mu}$ and $A_{\mu}$ notice that
if the mode is massless, then eq.\ (\ref{muv}) can be integrated once to find 
that $A_{\mu}' = \beta Z_{\mu}$. This allows us to
rewrite the equation for $Z_{\mu}$ in a decoupled form:
\beq
Z_{\mu}'' + \ell Z_{\mu}'  + \ell' Z_{\mu} + 2 \beta^2 Z_{\mu} =0.
\label{adec}
\eeq
It is already known that 
in the case of the six-dimensional Abelian vortex there exists a normalizable vector zero mode 
\cite{MG1,MG2} 
(see also \cite{ranshap}). In our case, recalling (\ref{uw2}), the zero mode takes the form 
\beq
A_{\mu } \sim \beta \int^{\rho} L d\rho' ,~~~~~~Z_{\mu} \sim L
\label{mode}
\eeq
which is the ground state solution of eqs.\ (\ref{muthperv}) and (\ref{muv}). 
This mode {\it is} normalizable since the canonical fields related to $A_{\mu}$ and $Z_{\mu}$ are 
simply the original ones multiplied by $\sqrt{L}$. The localization of the gauge mode 
related to $A_{\mu}$ was recently invoked \cite{MG1,MG2,ranshap} as a mechanism in order to localize gauge fields 
in the presence of thick Abelian strings in six-dimensions. 

The masslessness of the graviphoton mode is not a concern, as would be a massless scalar mode,
due to the fact that the graviphoton couples only to $T_{\mu\theta}$, which vanishes on the 
3-brane.  Since the graviphoton does not couple to standard model matter, it is not
phenomenologically constrained like a Brans-Dicke scalar. Moreover it does not seem a likely
candidate for explaining the tuning of the bulk geometry in response to a 3-brane tension,
since this is a deformation which looks like a purely scalar mode. 

\section{Discussion and conclusions}
We have made a comprehensive review of 6D cosmologies which contain at least one 3-brane
where the standard model could presumably be localized.   Emphasis was given to several
solutions in which the rate of expansion of the universe is either insensitive to the
tension of a 3-brane, eqs.\ (\ref{friedeq0}, \ref{friedeq1}), or else depends on $\tau_3$ in a way which
is at odds with 4D general relativity eq.\ (\ref{friedeq2}). This is despite the fact
that we are considering values of $\tau_3$ which are well below the scale of
compactification and the  mass scale of moduli which could account for a departure from
conventional gravity.  If the theories under consideration contained a massless radion,
for example, one would not be so surprised to see such behavior.  However both of the
models we have focused on do have stabilized moduli.  The AdS soliton model has only a
single radion among the fluctuations with azimuthal symmetry.  It is massless for
$\alpha=5$ and has an exponentially small negative (mass)$^2$ for $\alpha<5$ in the
absence of stabilization, but with the Goldberger-Wise mechanism it acquires a positive
(mass)$^2$ of order $m^4 k^{-2}\phi^2(\rho_m)e^{-2k\rho_m/5}\sim \lambda^2
e^{-2k\rho_m/5}$ \cite{BCCF}.  For realistic values of the hierarchy, this gives a
radion mass of order MeV, which is safe because the couplings of the radion to the
3-brane are suppressed by the 4D Planck scale.  And in the previous section we have
confirmed the stability of the unwarped ``football-shaped'' model by doing an exhaustive
study of its perturbation spectrum.

The tensions of 3-branes in two 6D theories fail to contribute to the Hubble rate in the
expected way at low energies, despite the fact that these theories have no additional massless
scalars that might explain a deviation from  4D general relativity.  Instead, the explanation
of this apparent failure of decoupling lies in the special way in which a 2D manifold's
internal geometry responds to the tension of a codimension two brane.  This can be seen from a
direct dimensional reduction of the 6D theory to four dimensions \cite{Luty}.  

Consider the metric (\ref{ansatz}) with a fluctuating 4D metric,
$g_{4,\mu\nu}(x_\mu)$.  When we integrate the 6D Einstein-Hilbert action over the coordinates
$y_i$ of the internal space, we obtain
\beqa
\label{dimred}
	{\cal L}_{4,\rm eff} &=& 
	\frac{1}{2\kappa_6^2}\sqrt{|g_4|} \int d^2 y\,\sqrt{g_2} M^{2}R_4 - V_{4,\rm eff}\\
\label{effpot}
	V_{4,\rm eff} &=& \int d^2 y\, M^4 \left(\sqrt{g_2}\left( -\frac{1}{2\kappa_6^2}
	R_2 + \Lambda_6 + \frac14 F^2 +\dots \right) + 
	\sum_i\tau_{3,i}\delta^{(2)}(y-y_i)  \right)
\eeqa
where $R_{n}$ is the $n$-dimensional curvature constructed from $g_{n,\mu\nu}$; $\tau_{3,i}$ is
the tension of the $i$th 3-brane, and $\dots$ denotes contributions from other sources
such as the bulk scalar field or a 4-brane.  In the unwarped solution, where 
$\tau_{3,1} = \tau_{3,2}\equiv\tau_3$,
there are {\it two} cancellations which make $R_4$ independent of $\tau_3$.  First, the
curvature $R_2$ has singular contributions that exactly cancel the delta function terms
explictly involving the tensions.  
One can see the crucial difference between codimension one and two branes by examining their
respective contributions to $R_2$ \cite{LMW}:
\beq
	{R_2\over\kappa_6^2} = 2\tau_3\, \delta^{(2)}(y) + \frac12 T_4\, \delta^{(1)}(y-y_4)
	+\hbox{nonsingular terms}
\eeq
The coefficient of the $2D$ delta function is such as to cancel the similar contribution to the
action from the 3-brane's stress-energy, whereas this cancellation is not exact for the 4-brane.
The second cancellation of the dependence on the 3-brane tension is this:
$\tau_3$ enters implicitly through the volume of the
extra dimension, $\sqrt{g_2(0)} = L'(0) = 1-\tau_3/2\pi$ which multiplies the entire nonsingular part of the
action, including the coefficient of $R_4$ in (\ref{dimred}), which
gives the 4D Planck mass:
\beq	
	M_p^2 = \frac{1}{\kappa_6^2}\int d^{\,2}\!y \sqrt{g_2}\, M^{2}(y)
\eeq
These two factors of $L'(0)$ cancel out of the Friedmann equation, $H^2 = V_{4,\rm
eff}/3M_p^2$, so that the Hubble rate is determined completely by bulk quantities, independent
of the brane tension.  Notice that one reason we can make this argument so simply in the
unwarped case is  the fact that the bulk solution changes only through a rescaling of $L$ as a
result of changing the brane tension.

When we try to make a similar argument to deduce the form of the Friedmann equation in the
warped solution, it is not straightforward, because $M$ and $L$ {\it do} change from their
static forms when the brane tension is changed, which induces a nonzero value of  $\Lambda_4$. 
It is still true that the delta function term  $\tau_{3}\delta^{(2)}(y)$ is cancelled by a
singular term of exactly the same magnitude in $\sqrt{g_2}R_2$.  But now there are additional
terms of order $\Lambda_4$ from the perturbation to the bulk part of the solution.  These did
not appear in the unwarped case.  The essential difference between the two models is that the
jump conditions at the 4-brane bring in a new source of dependence on $L$, hence the 3-brane
tension, which does not exist in the type 1  solution or in the unwarped type 3 solution. 
These observations were also made in \cite{Luty}.

The hope would be to solve the cosmological constant problem by using bulk supersymmetry to
ensure $\Lambda_4=0$, in a model where $\Lambda_4$ is independent of the 3-brane tension. Then
$\Lambda_4$ will be insensitive to all quantum corrections to the vacuum energy which arise if
the standard model is confined to the 3-brane. {However to describe realistic cosmology, one
needs to depart from pure de Sitter space and look for Friedmann-Robertson-Walker solutions.} 
In fact, it is not obvious that solutions to
the 6D Einstein equations exist when cosmological matter and radiation are added to the brane.
Let us suppose the line element has the rather general form
\beq
	ds^2 = -N^2 dt^2 + M^2 d\vec x^2 + B^2 d\rho^2 + L^2 d\theta^2
\eeq
where all the metric functions can depend on $t$ as well as $\rho$, 
and consider the terms in the Einstein equations which can have delta function singularities:
\beqa
	00&:&\quad 3\left({M''\over M} + {M'L'\over ML}\right)+ {L''\over L}  
	\,~~~~~~~~~~~~~~~~~~~~~~~\quad\sim\quad \epsilon B^2\delta^{(2)}(\rho)\\
	ii&:&\quad 2\left({M''\over M} + {M'L'\over ML}\right) 
	+ \left({N''\over N} +  {N'L'\over NL}\right) +{L''\over L} \quad\sim\quad 
	p B^2\delta^{(2)}(\rho)\\
	\theta\theta&:&\quad 3\left({M''\over M} +{M'L'\over ML}\right) 
	+ \left({N''\over N} +  {N'L'\over NL}\right)~~~~~~~\quad\sim\quad 0
\eeqa
In all the solutions we have considered, the energy density of the brane is given by
$\varepsilon=-p=\tau_3$, and $M'=N'=0$ at $\rho=0$, so that
there is no singular part in $M''$ or $N''$.  To obtain $\varepsilon\neq -p$, it is necessary to have
singular behavior not only in $L''/L$, but also in the other terms.  However $L''/L$ gives a 2D
delta function $\delta^{(2)}(\rho)\sim \delta(\rho)/\rho$ by virtue of $L$ vanishing like
$\rho$.  We can only get this kind of behavior from $M''/M$ if $M$ also vanishes, which is not
physically sensible since the standard model requires a nonvanishing metric.  

The only way we see to admit a general equation of state is to give the 3-brane a finite radius
$\rho_0$.  A solution to the equation $M''+M'/r = -\varepsilon M
\delta^{(2)}(\rho)$ can be found if $\delta^{(2)}(\rho)$ is a regularized delta function:
\beq
	M(\rho) = \left\{\begin{array}{ll} J_0(a\rho),& \rho<\rho_0\\
			 J_0(a\rho_0) + a\rho_0 {J_1(a\rho_0)}\ln{\rho_0\over\rho},& \rho>\rho_0
	\end{array}\right.
\eeq
where $a=\sqrt{\varepsilon/(\pi\rho_0^2)}$.  For branes of codimension one, the
interesting corrections to the Hubble rate come from terms like $(M'/M)^2$, which are
nonsingular and  have an unambiguous value proportional to $\varepsilon^2$.  For the
regularized codimension two brane, $(M'/M)^2 \sim (\varepsilon\rho/\rho_0^2)^2$ which
goes like $1/\rho_0^2$ when averaged over the brane.  Thus we expect to get results
which are sensitive to the internal structure of the brane, unlike the clean predictions
that came from the codimension one case.  Of course, unambiguous predictions could be
made in a specific theory, for example where the brane was a cosmic string defect coming
from an Abelian Higgs model.   Whether the interesting cosmological properties of
codimension two branes will survive after regularizing them requires further
investigation.

\bigskip
\noindent{\bf Acknowledgment.}  
We thank Daniel Chung, Rob Leigh and Dominik Schwarz for valuable discussions.  
J.C., J.D.\ and J.V.\ are supported in part by Canada's National Sciences and Engineering Research
Council.

\section*{Appendix}
\appendix

\section{Perturbation to AdS soliton from small $\Lambda_4$}
Let us write $E = E_0 + \delta E$, where $E_0 = az_0^2$ and
$\delta E = - b z_0^{6/5}$ is an energy shift relative to the static solution which allows us
to keep the same starting value $z_0$;  then $(E-U)$ can be written as
\beq
	E-U(z) =  a(z_0^2 - z^2) + b(z^{6/5} - z_0^{6/5})
\eeq
and the $b$ term can be consistently treated as a perturbation, even in the region $\rho\cong 0$, 
$z\cong z_0$.  The integral (\ref{rhoint}) gives
\beqa
	\rho &=& k^{-1}\ln\left({z\over z_0}+\sqrt{\left({z\over z_0}\right)^2 - 1}\right) -
	\delta\rho(z);\\
\label{deltarho}
	\delta\rho(z) &=& {b\over k^{3}} \int_{z_0}^z {z^{6/5}-z_0^{6/5} \over
	(z^2-z_0^2)^{3/2}} \, dz + O(b^2)\nonumber\\
	&\cong&  {b\, z_0^{4/5}\over k^3} \left\{\begin{array}{ll} 
	1.368 + \frac54\left({z_0\over z}\right)^{4/5} - \frac12\left({z_0\over z}\right)^{2}
	+ O( (\frac{z_0}{z})^{14/5}), & z\gg z_0\\
	0.85\,\epsilon^{1/2} - 0.18\,\epsilon^{3/2} + 0.062\,\epsilon^{5/2} + O(\epsilon^{7/2}),& \frac{z}{z_0} = 1+\epsilon \end{array}\right.
\eeqa
where now we define $k=|2a|^{1/2}$.
This can be inverted to find $z(\rho)$ and $M(\rho)$ to first order in $b$,

\section{Cosmology of codimension one branes in six dimensions}

In section 2, we mentioned that the case where $z'(0)\neq 0$ corresponds to having a 4-brane 
rather than a 3-brane at the origin, and that such a model is similar to RS in 5 dimensions.  
Here we will show this explicitly, following closely the formalism of \cite{CFCV}.

We write the metric as
\beqa
\label{metric}
ds^2 = -n^2(r,t)dt^2 + a^2(r,t)dx^2 + b^2(r,t)dr^2 + c^2(r,t) d\theta^2
\eeqa
and include a scalar to stabilize the bulk.  We will write the branes' stress energy 
tensors as:
\beqa
T^m_n &=& \delta(br) \diag(V_0+\rho_*,V_0-p_*,V_0-p_*,V_0-p_*,0,V_0-p_*^\theta)\nonumber\\
&&+\delta(b(r-R)) \diag(V_1+\rho,V_1-p,V_1-p,V_1-p,0,V_1-p^\theta)
\eeqa
where the perturbations $\rho$, $p$ have arbitrary equation of state.
We will expand the metric components around the static solution
\beqa
\label{expansion}
n(r,t)&=&\tilde a(r)e^{-N_1(r,t)};\quad a(r,t)=\tilde a(r) a_0 (t) e^{-A_1(r,t)}\nonumber\\
b(r,t)&=& 1 + B_1 (r,t);\quad c(r,t)=\tilde a(r) c_0(t) e^{-C_1(r,t)}\nonumber\\
\phi(r,t)&=&\phi_0(r)+\phi_1(r,t)
\eeqa
where $\tilde a$ and $\phi_0$ correspond to the static solutions and the other
terms are higher order in powers of $\rho$. 

\subsection{Boundary conditions}

The complications related to having a codimension-2 brane, and a 4-brane for which $T^0_0\neq T^{\theta}_{\theta}$ 
are not present in this model, since we are simply dealing with a space cut off by two codimension-1 branes.  
We simply impose $S_1/Z_2$ symmetry, so that the boundary conditions will be 
\beqa
\label{bcR}
\left.\left(3\frac{a'}{a}+\frac{c'}{c}\right)\right|_{y=R} &=& \frac{b(R)\kappa^2}{2}\left.{T^0}_0\right|_{y=R}\nonumber\\
\left.\left(2\frac{a'}{a}+\frac{c'}{c}+\frac{n'}{n}\right)\right|_{y=R} &=& \frac{b(R)\kappa^2}{2}\left.{T^i}_i\right|_{y=R}\nonumber\\
\left.\left(3\frac{a'}{a}+\frac{n'}{n}\right)\right|_{y=R} &=& \frac{b(R)\kappa^2}{2}\left.{T^5}_5\right|_{y=R}
\eeqa

\beqa
\label{bc0}
\left.\left(3\frac{a'}{a}+\frac{c'}{c}\right)\right|_{y=0} &=& -\frac{b(0)\kappa^2}{2}\left.{T^0}_0\right|_{y=0}\nonumber\\
\left.\left(2\frac{a'}{a}+\frac{c'}{c}+\frac{n'}{n}\right)\right|_{y=0} &=& -\frac{b(0)\kappa^2}{2}\left.{T^i}_i\right|_{y=0}\nonumber\\
\left.\left(3\frac{a'}{a}+\frac{n'}{n}\right)\right|_{y=0} &=& -\frac{b(0)\kappa^2}{2}\left.{T^5}_5\right|_{y=0}.
\eeqa
We must also consider the boundary conditions for the scalar field equation, which are
\beqa
\frac{2}{b}\phi'|_{r=0} &=&V_0'\nonumber\\
\frac{2}{b}\phi'|_{r=R} &=&V_1'
\eeqa

\subsection{Order ($\rho^0$)}

The static solutions must solve the Einstein and scalar field equations expanded to zeroth order in $\rho$. 
These can be written as
\beqa
\label{order0EE}
\frac{\partial}{\partial \phi_0} V(\phi_0)&=&\phi_0''+\phi_0'5\frac{\tilde a'}{\tilde a}\nonumber\\
V(\phi_0)&=&-\Lambda+\frac{1}{2}\phi_0'^2-10\kappa^{-2}\left(\frac{\tilde a'}{\tilde a}\right)^2\nonumber\\
-4\frac{\tilde a''}{\tilde a}&=&\kappa^2(V(\phi_0)+\frac{1}{2}\phi_0'^2+\Lambda)+6\left(\frac{\tilde a'}{\tilde a}\right)^2
\eeqa
Using the superpotential method \cite{dewolfe}, one can find an exact solution
\beqa
\tilde a(r) &=&\tilde c(r) = e^{-A_0(r)}\nonumber\\
A_0(r)&=& \frac{2}{5}kr+\frac{\kappa^2 v_0^2}{16}\left(e^{-2\epsilon kr}-1\right)\nonumber\\
\phi_0&=&v_0 e^{-\epsilon kr}\nonumber\\
V(\phi_0)&=&\epsilon k^2 \phi_0^2 \left(1+\frac{\epsilon}{2}-\frac{5}{32}\kappa^2 \epsilon \phi_0^2\right)
\eeqa
The boundary conditions at lowest order in $\rho$ impose:
\beqa
V_0&=&-\frac{10\Lambda}{k}\left(1-\frac{5}{16}\kappa^2v_0^2\epsilon\right)\nonumber\\
V_1&=&\frac{10\Lambda}{k}\left(1-\frac{5}{16}\kappa^2v_0^2\epsilon e^{-2\epsilon kR}\right)
\eeqa
where
\beqa
\Lambda = -\frac{8k^2}{5\kappa^2}.
\eeqa
It is clear that at the static level, this model is completely analogous to Randall-Sundrum.

\subsection{Order $\rho$}

We now look at the perturbed equations.  We first define the following variables:
\beqa
\label{newvars}
\Psi_1&=&-\left(3A'_1+C'_1+4B_1\frac{\tilde a'}{\tilde a}\right)+\kappa^2 \phi_0'\phi_1 \nonumber\\
\Upsilon_1&=& N'_1 -A'_1\nonumber\\
\xi_1&=&-\left(3A'_1+N'_1+4B_1\frac{\tilde a'}{\tilde a}\right)+\kappa^2 \phi_0'\phi_1\nonumber\\
\zeta_1&=&\phi_0''\phi_1-\phi_0'\phi_1'+\phi_0'^2 B_1
\eeqa
which appear naturally in the boundary conditions. It is simple to check that these are invariant 
under the following gauge transformations:
\beqa
y&=&\bar y +f(\bar y) ;\quad f(0) = f(1) = 0\nonumber\\
A_1 &\rightarrow& A_1 - \frac{\tilde a'}{\tilde a}f ;\quad N_1 \rightarrow N_1 - \frac{\tilde a'}{\tilde a}f\nonumber\\
B_1 &\rightarrow& B_1 + f' ;\quad \phi_1 \rightarrow \phi_1 + \phi_0' f\nonumber\\
C_1 &\rightarrow& C_1 - \frac{\tilde a'}{\tilde a}f. 
\eeqa

These new variables obey the boundary conditions
\beqa
\label{order1bc}
\left.\Psi_1\right|_{y=R}&=&\frac{\kappa^2}{2}\rho;\quad\left.\Psi_1\right|_{y=0}=-\frac{\kappa^2}{2}\rho_*\nonumber\\
\left.\Upsilon_1\right|_{y=R}&=& \frac{\kappa^2}{2}\left(\rho+p\right);\quad \left.\Upsilon_1\right|_{y=0}= -\frac{\kappa^2}{2}\left(\rho_*+p_*\right)\nonumber\\
\left.\xi_1\right|_{y=R}&=&-\frac{\kappa^2}{2}p^\theta;\quad \left.\xi_1\right|_{y=0}=\frac{\kappa^2}{2}p_*^\theta
\eeqa
Since we still haven't fixed a gauge, we will now choose to work in the stiff potential limit,
i.e. $\phi_1\equiv 0$ \cite{CFCV}.  This allows us 
to write the first order in $\rho$ terms of the Einstein equations as :

\beqa
3 e^{2A_0}\left[\left(\frac{\dot a_0}{a_0}\right)^2+\frac{\dot a_0}{a_0}\frac{\dot c_0}{c_0}\right]_1&=&\Psi_1' -5 A_0'\Psi_1 \\
e^{2A_0}\left[2\left(\left(\frac{\dot a_0}{a_0}\right)^2-\frac{\ddot a_0}{a_0}\right) + \frac{\dot a_0}{a_0}\frac{\dot c_0}{c_0} - \frac{\ddot c_0}{c_0}\right]_1&=&\Upsilon'_1 -5A_0' \Upsilon_1\\
3e^{2A_0}\left[\frac{\ddot a_0}{a_0} + \left(\frac{\dot a_0}{a_0}\right)^2\right]_1&=&\xi'_1-5A_0'\xi_1\\
3e^{2A_0}\left[\frac{\ddot a_0}{a_0} + \left(\frac{\dot a_0}{a_0}\right)^2 + \frac{\dot a_0}{a_0}\frac{\dot c_0}{c_0} + \frac{1}{3}\frac{\ddot c_0}{c_0}\right]_1&=&-A_0'\left(\xi_1+4\Psi_1 -3\Upsilon_1\right) + \kappa^2\phi_0'^2 B_1\\
\dot\Psi_1+3\left(\frac{\dot a_0}{a_0}\right)_{1/2}\Upsilon_1+\left(\frac{\dot c_0}{c_0}\right)_{1/2}(\Psi_1-\xi_1)&=&0
\eeqa

 The solutions are given by:
\beqa
\Psi_1(r) &=& e^{5A_0(r)}\left[\Psi|_{r=0}+3\left[\left(\frac{\dot a_0}{a_0}\right)^2+\frac{\dot a_0}{a_0}\frac{\dot c_0}{c_0}\right]_1 \int^r_0 e^{-3A_0(r)}dr\right]\nonumber\\
\Upsilon_1(r) &=& e^{5A_0(r)}\left[\Upsilon|_{r=0}+\left[2\left(\left(\frac{\dot a_0}{a_0}\right)^2-\frac{\ddot a_0}{a_0}\right) + \frac{\dot a_0}{a_0}\frac{\dot c_0}{c_0} - \frac{\ddot c_0}{c_0}\right]_1 \int^r_0 e^{-3A_0(r)}dr\right]\nonumber\\
\xi_1(r) &=& e^{5A_0(r)}\left[\xi|_{r=0}+3\left[\frac{\ddot a_0}{a_0} + \left(\frac{\dot a_0}{a_0}\right)^2\right]_1 \int^r_0 e^{-3A_0(r)}dr\right]
\eeqa

The Friedmann equations are then, using $8\pi G \equiv\frac{\kappa^2}{2\int^R_0 e^{-3A_0(r)}dr}$,
\beqa
\left(\frac{\dot a_0}{a_0}\right)^2&=&\frac{8\pi G}{3}\left[\rho_*+e^{-5A_0(R)} \rho \right]-\frac{\dot a_0}{a_0}\frac{\dot c_0}{c_0}\nonumber\\
\frac{\ddot a_0}{a_0}-\left(\frac{\dot a_0}{a_0}\right)^2 &=& -4\pi G\left[\rho_*+ p_*+e^{-5A_0(R)}(\rho+ p)\right]+\frac{1}{2}\frac{\dot a_0}{a_0}\frac{\dot c_0}{c_0}-\frac{1}{2}\frac{\ddot c_0}{c_0}\nonumber\\
\frac{\ddot a_0}{a_0}+\left(\frac{\dot a_0}{a_0}\right)^2 &=& -\frac{8\pi G}{3}\left[(p_*^\theta+e^{-5A_0(R)}p^\theta\right]
\eeqa
We can combine these to write:
\beqa
\label{ceqn}
3\frac{\dot a_0}{a_0}\frac{\dot c_0}{c_0}+\frac{\ddot c_0}{c_0}&=&\frac{8\pi G}{3}\left[\rho_*-3 p_*+ e^{-5A_0(R)}(\rho -3 p)+2\left(p_*^\theta+e^{-5A_0(R)}p^\theta\right)\right]
\eeqa
We can then use the fourth Einstein equation to solve for $B_1$:
\beqa
B_1&=&\frac{1}{6\phi_0'^2}\left[\left(\rho_* -3 p_* -p_*^\theta\right)\left(F(r)-3A_0'e^{5A_0}\right)+e^{-5A_0(R)}\left(\rho -3p -p^\theta\right)F(r)\right]\nonumber\\
\eeqa
where we have defined
\beqa
F=\frac{1}{\int_0^R e^{-3A_0(r)}dr}\left[e^{2A_0}+3A_0'e^{5A_0}\int_0^r e^{-3A_0(r)}dr\right]
\eeqa

These results are very similar to those one gets in the Randall-Sundrum scenarion.  There are however 
some differences.  First, $\rho$ and $p$ denote $(4+1)$-dimensional quantities, since the branes 
in this model correspond to 4-branes.  If the compact dimension is small however, there will be 
no light Kaluza-Klein modes, and one expects to recover standard cosmology at low energies.

Another difference is the appearance of terms involving the time variation of the
compact dimension's  scale factor. Such terms are observationally constrained to be
quite small, so that they should not lead to appreciable effects on late-time cosmology
once a  stabilization mechanism for the size of the angular dimension is included
\cite{CV}.


\end{document}